\documentclass[aps,prb,showpacs,twocolumn,floats,epsfig,pdflatex, superscriptaddress,longbibliography]{revtex4-2}
\usepackage{amssymb}
\usepackage{amsbsy}
\usepackage{amsmath}
\usepackage{epsfig}
\usepackage{amsfonts}
\usepackage{graphicx}
\usepackage{amssymb}
\usepackage{upgreek}
\usepackage{multirow}
\usepackage{dsfont}
\usepackage{float}
\usepackage{bm}
\usepackage{xcolor, comment}
\usepackage{hyperref}
\usepackage{tikz}

\begin{document}

\title {Raman tensor for two-dimensional massive Dirac fermions}

\author{Selçuk Parlak}
\affiliation{D\'epartement de Physique, Institut Quantique and Regroupement Qu\'eb\'ecois sur les Mat\'eriaux de Pointe, Universit\'e de Sherbrooke, Sherbrooke, Qu\'ebec, Canada J1K 2R1}
\author{Ion Garate}
\affiliation{D\'epartement de Physique, Institut Quantique and Regroupement Qu\'eb\'ecois sur les Mat\'eriaux de Pointe, Universit\'e de Sherbrooke, Sherbrooke, Qu\'ebec, Canada J1K 2R1}
\date{\today}

\begin{abstract}
Raman spectroscopy is a valuable characterization tool for two-dimensional materials. 
Starting from model Hamiltonians for Chern insulators
and magnetized monolayers of transition metal dichalcogenides, we theoretically predict two unconventional features of Raman spectroscopy.
First, a selection rule emerges in the Raman
tensor when the incident and scattered photons are circularly polarized. This rule generalizes the well-known
valley selection rule of optical conductivity in Dirac insulators.
Second, for an electronic model with a single massive Dirac fermion, the phase diﬀerence between Raman tensor elements is quantized to $\pm\pi/2$ for
any frequency of the incident light. The quantization is robust under perturbations and the sign of the phase diﬀerence is reversed when the mass term of the Hamiltonian is inverted. 
\end{abstract}

\maketitle
\section{Introduction} 

Twenty years after the synthesis of graphene \cite{novoselov2004electric}, two dimensional (2D) Dirac materials continue to be an active subject of research in condensed matter physics and electrical engineering, both for their fundamentally interesting properties and for their potentially useful applications (for recent reviews, see e.g. Refs. \cite{fan2023two, lin2023recent}). \par
Among the variety of experimental techniques used to probe 2D Dirac materials, Raman spectroscopy stands out as a noninvasive tool that characterizes their structural, vibrational, magnetic and elastic properties \cite{malard2009raman, saito2016raman, wu2018raman, cong2020application}.
Historically, Raman scattering in graphene monolayers and multilayers  \cite{ferrari2013raman,bonaccorso2012production} gave information about the number and  placement of layers, and about defects generated during production \cite{ferrari2006raman,bonaccorso2012production,ferrari2013raman}. More recent developments on the synthesis of 2D transition metal dichalcogenides, with sizable band gaps close to visible light frequencies, have steered the attention of Raman spectroscopists to 2D Dirac materials hosting massive Dirac fermions \cite{zhang2015phonon}.

In polarized Raman spectroscopy \cite{loudon1963theory, cardona2005resonance}, photons of frequency $\omega_1$, wave vector ${\bf k}_1$ and polarization $\hat{\bf e}_1$ scatter off the sample and are detected with frequency $\omega_2$, wave vector ${\bf k}_2$ and polarization $\hat{\bf e}_2$. 
The intensity of the scattered light from a phonon mode $\lambda$ is given by
\begin{equation}
\label{eq:rint}
 I_\lambda \propto | \hat{\bf e}_1^\dagger\cdot {\bf R}_\lambda\cdot\hat{\bf e}_2|^2,
\end{equation}
where ${\bf R}_\lambda$ is the Raman tensor.  In general, the Raman tensor elements are functions of $\lambda$, $\omega_1$, $\omega_2$, the details of the electronic band structure and the electronic decay rate $\eta$.

 Polarized Raman measurements have been used in 2D materials e.g. to identify individual phonon modes, to determine the crystallographic orientation, and to analyze the phases of the Raman tensor elements \cite{kim2020polarized,pimenta2021polarized}.

In parallel to the experimental Raman measurements in 2D Dirac materials, theorists have elucidated  the ``transition sliding'' in graphene \cite{heller2016theory}, explored the breakdown of the adiabatic Born-Oppenheimer approximation in the Raman G peak of graphene \cite{pisana2007breakdown}, explained Raman selection rules \cite{huang2022new} and the influence of external perturbations  \cite{kukucska2017theoretical} in MoS$_2$ monolayers, detected Raman signatures of massive Dirac fermions in bilayer graphene \cite{graziotto2024infrared} or computed the Raman spectra for hundreds of monolayers \cite{taghizadeh2020library} to name but a few representative examples.

To date, nearly all of the theoretical work on Raman scattering in 2D Dirac materials has been based on  {\em ab-initio} electronic structure calculations.   Though these methods have the merit of often offering a good agreement with the measured Raman spectra,
they rarely provide simple analytical insights into how the Raman tensor inherits traits of the low-energy Dirac-like electronic dispersion. 
Exceptionally, Ref. \cite{basko2009calculation} has used a low-energy Dirac Hamiltonian to obtain analytical expressions for the Raman tensor associated to in-plane phonons in graphene.
Yet, graphene is special in that it hosts massless Dirac fermions. All insulating 2D Dirac materials have a nonzero Dirac mass, and indeed that mass is crucial to realize topological phases \cite{ren2016topological}.

To our knowledge, there are no model-Hamiltonian calculations of the Raman tensor for massive 2D Dirac fermion systems.
In Ref.~\cite{sasaki2018phonon}, the influence of massive Dirac fermions on phonon dispersion has been theoretically explored.
More recently, Raman signatures of the electronic valley Chern number, proportional to the phonon wave vector, have been predicted in BaMnSb$_2$ \cite{hu2021phonon, parlak2023detection}.
None of these works carry out a microscopic theory calculation of the Raman tensor.
As a result, fundamental questions such as ``is the Raman tensor for zone-center phonons sensitive to the sign change of the Dirac mass?'', which have a bearing in the characterization of topological phases, have not been addressed.  The purpose of our work is to explore such questions.

Our main results can be found in Sec. \ref{sec:onevalley}, where we calculate the contribution from a single 2D massive Dirac fermion to ${\bf R}_\lambda$ using the standard perturbative approach \cite{loudon1963theory, basko2009calculation}. 
In this model, ${\bf R}_\lambda$ is a $2\times 2$ matrix with elements $R_{ij}$ ($i,j\in\{x,y\})$, with $xy$ being the plane of the 2D material.
We focus on zero temperature, in which only the Stokes (phonon emission) processes occur. Accordingly, $\omega_2=\omega_1-\omega_{\bf q}$ and ${\bf k}_2= {\bf k}_1-{\bf q}$, with $\omega_{\bf q}$ and ${\bf q}$ being the frequency and wave vector of the photoexcited phonon 
\footnote{The main results of our manuscript hold more generally to the case in which $\omega_1-\omega_2$ is not necessarily equal to $\omega_0$, provided that $|\omega_1-\omega_2|\ll {\rm min}(\omega_1, \omega_2)$.}. 
We limit ourselves to long-wavelength optical phonons, for which we set ${\bf q}=0$ and $\omega_{\bf q}= \omega_0$. We also suppose that the Fermi energy is inside the energy gap of the insulator. 

Because a single 2D Dirac fermion breaks time-reversal symmetry but preserves rotational symmetry in the $xy$ plane, we find $R_{xx}=R_{yy}$ and $R_{xy}=-R_{yx}$ for out-of-plane lattice vibrations that modulate the Dirac mass. While the analytical expressions for $R_{xx}$ and $R_{xy}$ are cumbersome, the  ratio $R_{xy}/R_{xx}$ is unexpectedly very simple. This simplicity leads to two properties.  First, there is a Raman selection rule under illumination by circularly polarized light, which can be regarded as a generalization of the optical valley selection rule \cite{wang2008valley,zeng2012valley,mak2018light}. Second, the phase difference between $R_{xx}$ and $R_{xy}$ is quantized to $\pm \pi/2$, independently of the frequency of the light. Moreover, the sign of the phase difference is set by the signs of the Dirac mass and Dirac velocities.

In Sec. \ref{sec:generalizations}, we extend our results to magnetized 2D crystals hosting multiple Dirac fermions (two in Sec.~\ref{sec:two} and four in Sec.~\ref{sec:four}). 
The motivation here is to bring our model calculation closer to the electronic structure of transition metal dichalcogenide monolayers with broken time-reversal symmetry. 
This generalization comes at the expense of losing some of the simplicity from the results of Sec.~\ref{sec:onevalley}. Nevertheless, the quantization of the phase difference and the Raman selection rule remain robust (the former when the frequency of the incident light is smaller than the energy gap of the insulator).
 We are not aware of any experiment exploring our predicted selection rule and phase difference in magnetized 2D Dirac materials.

Finally, Sec. \ref{sec:conc} summarizes the main findings and presents the conclusions. The appendices collect some technical details about the calculation of the Raman tensor.
We take $\hbar\equiv 1$ throughout the paper.

\section{One Dirac fermion}
\label{sec:onevalley}


In this section, we analyze the contribution from a single 2D massive Dirac fermion to the Raman tensor.

 The Hamiltonian of interest is
\begin{equation}
{\cal H}={\cal H}_{\text{e}}+{\cal H}_{\text{e-pt}}+{\cal H}_{\text{e-pn}},
\end{equation}
where ${\cal H}_\text{e}, {\cal H}_{\text{e-pt}}$ and ${\cal H}_{\text{e-pn}}$ are the electron, electron-photon and electron-phonon Hamiltonians respectively.
First, the electronic part reads
\begin{equation}
\label{eq:He}
{\cal H}_{\text{e}}=\sum_{\bf k} \Psi_{\bf k}^\dag \left[(v_x k_x \sigma_y- v_y k_y \sigma_x)+m \sigma_z\right]\Psi_{\bf k},
\end{equation}
 where ${\bf k}=(k_x, k_y)$ is the electronic wave vector, $\Psi_{\bf k}^\dagger$ and $\Psi_{\bf k}$ are electronic creation and annihilation operators, $v_{x,y}$ is the Dirac velocity in the $x$ or $y$ direction, $\sigma_{x,y,z}$ are Pauli matrices, and $m$ is the Dirac mass. 
 In this work, we assume that the Fermi energy is in the gap, and we take the zero temperature limit.
 
Second, the electron-photon interaction is given by 
\begin{equation}
{\cal H}_{\text{e-pt}}=-e \sum_{{\bf k}} \Psi_{\bf k}^\dag  (v_x A_x  \sigma_y-v_y A_y   \sigma_x )\Psi_{\bf k},
\label{eq:electronpt}
\end{equation}
where $e$ is the electron's charge and $A_{x,y}$ are the $x-$ and $y-$components of the electromagnetic vector potential, whose wave vector we neglect. The $z$ component of the vector potential does not couple to electrons in our idealized model, where the electronic velocity is confined to the $xy$ plane.  Consequently, the Raman tensor will turn out to be a $2\times 2$ matrix. The ratio $A_x/A_y$ is determined by the polarization of the light (incident or scattered).


Third, 
the coupling between the 2D Dirac fermions and a nondegenerate phonon mode $\lambda$ of vanishing wave vector can be written as
\begin{equation}
{\cal H}_{\text{e-pn}}=\sum_{{\bf k}}\Psi_{\bf k}^\dag \left(\sum_i g_i \sigma_i u_\lambda\right) \Psi_{\bf k},
\label{eq:electronpn1}
\end{equation}
where $u_\lambda$ is the normal mode coordinate, $i\in\{0,x,y,z\}$, $\sigma_0$ is the identity matrix and $g_i$ are real numbers denoting the electron-phonon coupling.
The couplings $g_0$ and $g_z$ describe the phonon-induced modulations of the chemical potential and the Dirac mass, respectively, while $g_x$ and $g_y$ describe the phonon-induced shift of the energy  bands in ${\bf k}$-space. 
For simplicity, we neglect the dependence of $g_i$ in  ${\bf k}$.
If the phonon mode $\lambda$ is degenerate, Eq.~(\ref{eq:electronpn1}) is generalized to
\begin{equation}
{\cal H}_{\text{e-pn}}=\sum_{{\bf k}}\Psi_{\bf k}^\dag  \left(\sum_{i,j}g_{i j}\sigma_i u_{\lambda,j} \right)\Psi_{\bf k},
\label{eq:electronpn2}
\end{equation}
where $i\in\{0,x,y,z\}$, $j\in\{1,...,n_d\}$ and $n_d$ is the degeneracy of the phonon.
Below, we will be concerned with non-degenerate phonon modes.

The Dirac mass term in ${\cal H}_e$ breaks both the parity symmetry ${\cal P}$ and the time reversal symmetry ${\cal T}$, because $\sigma_x h_{\text{e}}(k_x,k_y,m) \sigma_x\neq h_{\text{e}}(-k_x,k_y,m)$ and $\sigma_y h^*_{\text{e}}({\bf k}, m) \sigma_y\neq h_{\text{e}}(-{\bf k}, m)$, respectively. Here, $h_{\rm e}({\bf k}, m)\equiv k_x\sigma_y-k_y \sigma_x+m\sigma_z$. In ${\cal H}_{\rm e-pn}$, having a ${\bf k}-$independent $g_i$ requires broken ${\cal T}$ ($u_\lambda$ being invariant under time-reversal).


The electronic Hamiltonian has continuous rotational symmetry around the $z$-axis (this is the case even when $v_x\neq v_y$, as can be seen by rescaling the wave vectors).
In Ref.~\cite{basko2009calculation}, it has been stated in the context of graphene that the Raman tensor should vanish when ${\cal H}_e$ has continuous rotational symmetry around the $z$ axis.
Yet, Ref.~\cite{basko2009calculation} considers only in-plane lattice vibrations. 
For out-of-plane lattice vibrations, the Raman tensor can be nonzero even though  ${\cal H}_e$ has continuous rotational symmetry around the $z$ axis. 
To see this, let $O$ be a symmetry operation of the system.
If $O$ leaves $u_\lambda$ invariant (as is the case for out-of-plane phonons under any rotation around the $z$ axis), then the Raman tensor obeys the following relation dictated by Neumann's principle:
\begin{equation}
\label{eq:neumann1}
R_{i j \lambda} = \sum_{l,m} O_{i l} O_{j m} R_{l m \lambda},
\end{equation}
where $i,j,l,m\in\{x,y\}$ and  $O_{i l}$ is the $(i, l)$ element of the matrix associated to a symmetry operation.
Substituting $O_{xx} = O_{yy} = \cos\theta$ and $O_{xy}=-O_{yx} = -\sin\theta$ for arbitrary $\theta$ in Eq.~(\ref{eq:neumann1}), we obtain
\begin{equation}
\label{eq:R_lambda}
{\bf R}_\lambda = \left(\begin{array}{cc} R_{xx} & R_{xy} \\ -R_{xy} & R_{xx}\end{array}\right),
\end{equation}
without any requirement for the vanishing of the Raman tensor elements.
For notational brevity, we have omitted (and we will hereafter omit) the phonon label $\lambda$ from the elements of the Raman tensor in Eq.~(\ref{eq:R_lambda}).
It is important to mention that the conditions $R_{xx}=R_{yy}$ and $R_{xy}=-R_{yx}$ do not necessarily require continuous rotational symmetry; they can also arise from {\em discrete} rotational symmetry, such as a $C_3$ axis along $z$.
\begin{figure}[t]
\begin{center}
   \includegraphics[width=1\columnwidth]{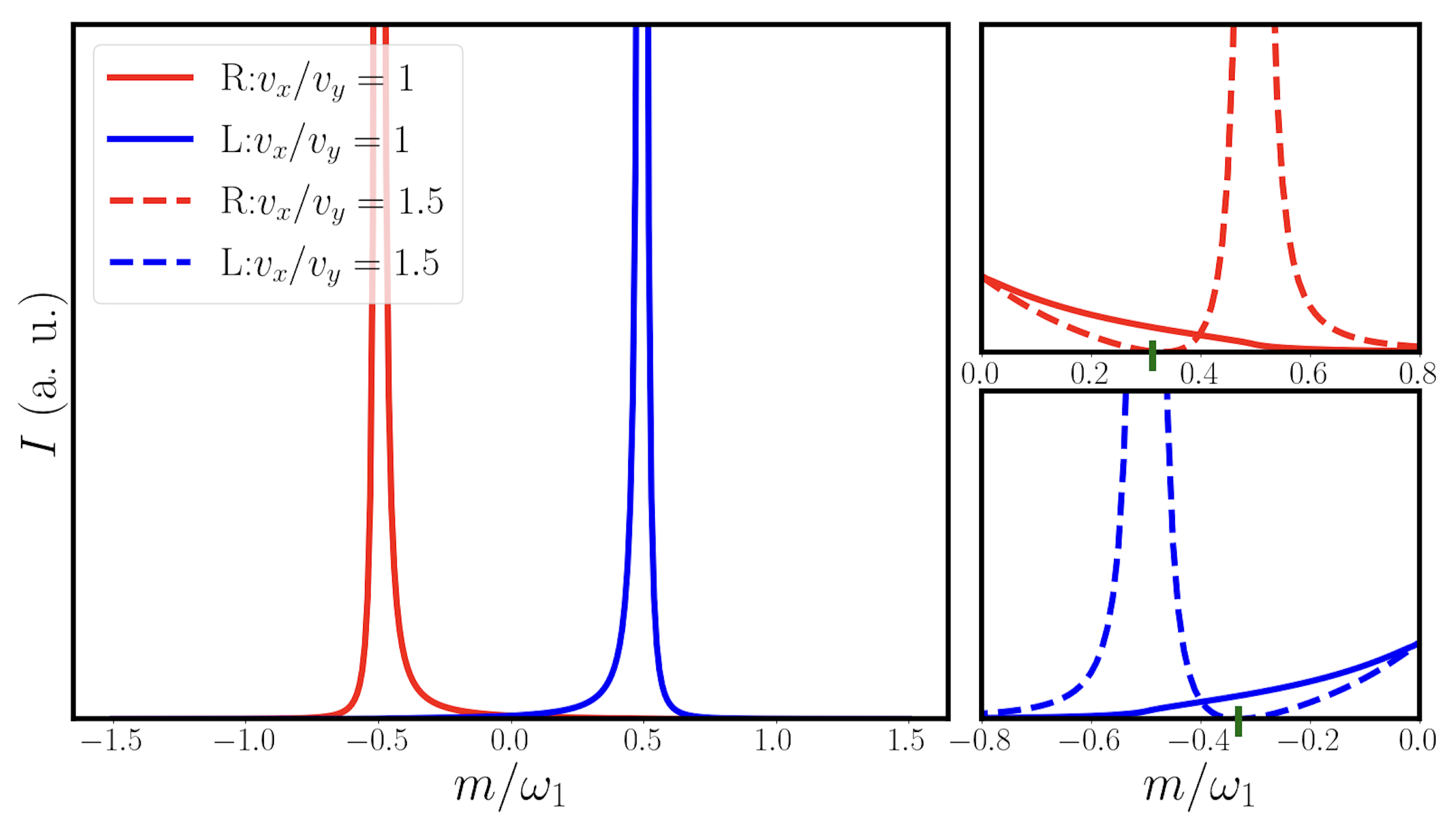}                            
\caption{Selection rule in the Raman intensity as a function of the Dirac mass $m$, for a fixed incident light frequency $\omega_1$. 
The electronic decay rate and the phonon frequency are $\eta=0.01 \omega_1$ and $\omega_0=0.01 \omega_1$, respectively.
The incident and detected lights have the same circular polarization ($R$ and $L$ for right- and left-circular polarization, respectively). 
The solid and dashed  lines correspond to isotropic ($v_x=v_y$) and anisotropic  ($v_x\neq v_y$) Fermi velocities, respectively.
For non-circularly polarized light, the resonant Raman scattering takes place at $m/\omega_1=\pm 0.5$. Yet, when the light is circularly polarized and $v_x=v_y$ (left panel), one of the resonances turns into  an extinction (at $m/\omega_1= 0.5$ for $R$ light and $m/\omega_1= -0.5$ for $L$ light). {\em Right panels:} When the light is circularly polarized and $v_x\neq v_y$, resonances at $m/\omega_1=\pm 0.5$ are both present, but the Raman intensity has a zero either at $m/\omega_1 = 0.5 |v_y/v_x|$ (for $R$ light) or at $m/\omega_1 = -0.5 |v_y/v_x|$ (for $L$ light), as represented by green vertical lines.} 
 \label{fig:selection}
  \end{center}
\end{figure} 
Another consequence of rotational symmetry around the $z-$axis is that, for out-of-plane vibrations, $g_x=g_y=0$ (here we use Eq.~(\ref{eq:electronpn1}), since out-of-plane phonons are nondegenerate).
This follows from the requirement 
\begin{equation}
\label{eq:sym}
{\cal C}_\theta^{-1} {\cal H}_{\rm e-pn} {\cal C}_\theta= {\cal H}_{\rm e-pn},
\end{equation}
 where ${\cal C}_\theta$ is a symmetry operator for a rotation of an angle $\theta$ around the $z$ axis.
Substituting 
\begin{align}
&{\cal C}_\theta^{-1} u_\lambda{\cal C}_\theta = u_\lambda\notag\\
&{\cal C}_\theta^{-1} \Psi^\dagger_{\bf k}{\cal C}_\theta = \Psi^\dagger_{{\bf k}'} \exp(-i \sigma_z \theta/2)
\end{align}
with $k_x'=k_x\cos\theta-k_y\sin\theta$ and $k_y'=k_y\cos\theta+k_x\sin\theta$, and recalling that $g_i$ are assumed to be independent of ${\bf k}$, Eq.~(\ref{eq:sym}) yields
\begin{align}
\label{eq:gxy}
g_x-i g_y &= (g_x-i g_y) \exp(i \theta)\notag\\
g_x+i g_y &= (g_x+i g_y) \exp(-i \theta).
\end{align}
For any $\theta\neq 0$, Eq.~(\ref{eq:gxy}) results in $g_x=g_y=0$.
Therefore, when it comes to out-of-plane lattice vibrations in a 2D Dirac fermion system with rotational symmetry, only phonons that produce Dirac mass or chemical potential modulations can couple to electrons (with couplings $g_z$ and $g_0$, respectively). 
Furthermore, from inspection of the microscopic theory expressions for the Raman tensor elements in Appendix \ref{ap:main}, it is clear that $g_0$ makes no contribution. 
Thus, we end up with $R_{xx}\propto g_z$ and $R_{xy}\propto g_z$. The linearity of the Raman tensor elements on the electron-phonon coupling is a consequence of the fact that we limit ourselves to one-phonon processes.
Also, we hereafter treat $g_z$ as a constant independent of electronic band parameters, such as $m$; relaxing this assumption would not influence the main results below. 
\begin{figure}[t]
\begin{center}
   \includegraphics[width=1\columnwidth]{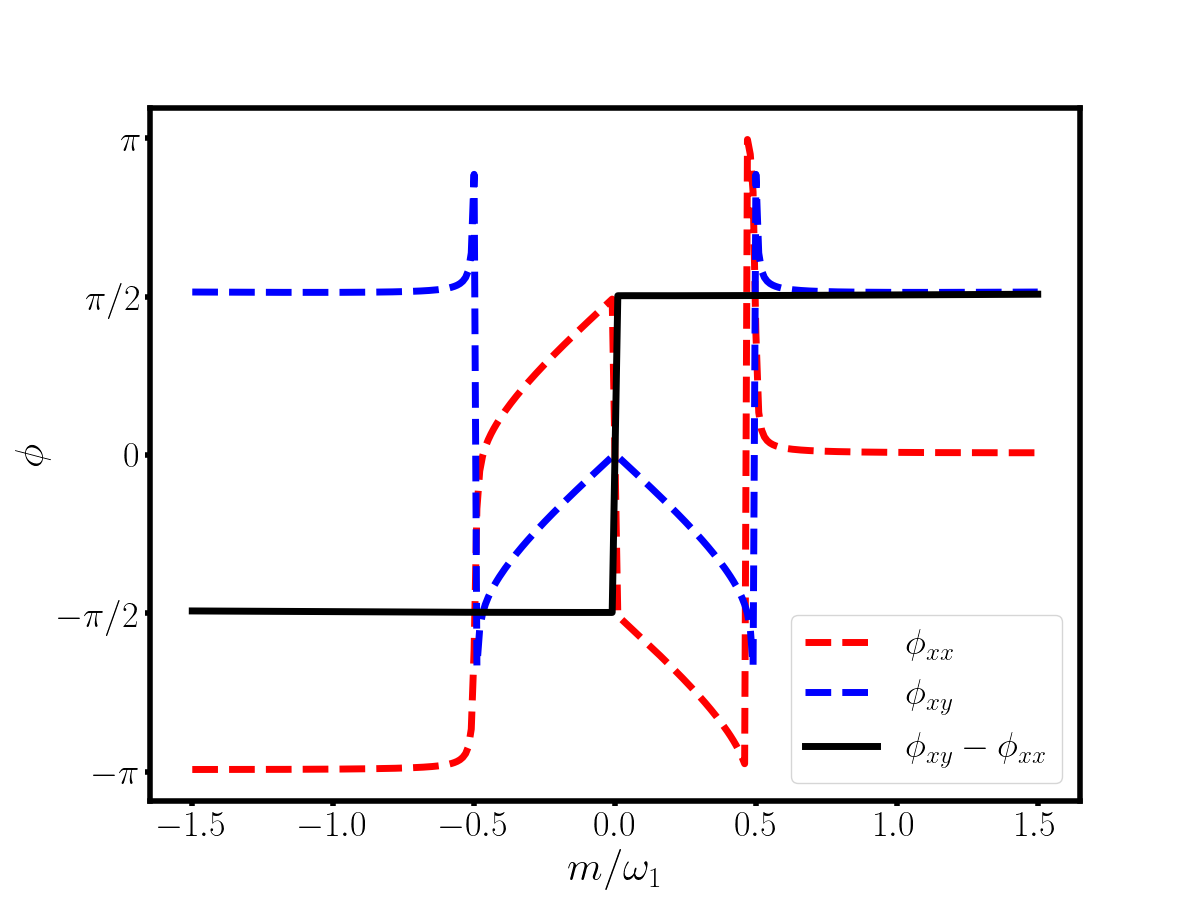}                            
\caption{Individual phases of the Raman tensor elements $R_{xx}$ and $R_{xy}$, and their difference, as a function of the ratio between the Dirac mass $m$ and the incident photon frequency $\omega_1$. Resonant Raman scattering takes place when $m/\omega_1 = \pm 0.5$. The electronic decay rate and the phonon frequency are taken as $\eta=0.01 \omega_1$ and $\omega_0 = 0.01, \omega_1$, respectively. The Fermi velocity is considered to be isotropic ($v_x=v_y$). The electron-phonon coupling $g_z$ is assumed to be independent of the mass (this assumption has no effect in $\phi_{xx}-\phi_{xy}$).}
 \label{fig:rxxrxy}
  \end{center}
\end{figure} 
According to Eq. (\ref{eq:R_lambda}), $R_{xy}\neq 0$ implies an antisymmetric Raman tensor.
At first glance, this finding is at odds with the common statement that Raman tensors are symmetric ($R_{ij}=R_{ji}$) except close to resonance 
\cite{loudon1963theory}.
Yet, in the argument of Ref.  \cite{loudon1963theory}, the electronic wave functions are assumed to be real, which requires
time-reversal symmetry, and the electronic decay rate $\eta$ is likewise neglected.
Those conditions do not apply to our model, where either $m\neq 0$ or $g_z\neq 0$ signifies broken time-reversal symmetry and $\eta$ plays a role.

Starting from the expressions in Appendix \ref{ap:main}, it is straightforward to derive the more general relation 
\begin{equation}
\label{eq:exc}
R_{ij}(\eta)=R_{ji}^*(-\eta),
\end{equation}
which is compatible with having an antisymmetric Raman tensor.
Equation \ref{eq:exc}) is valid far from resonance (i.e., when $\omega_1\pm \omega_0$ is not close to the energy gap of the insulator $2|m|$). 
All the parameters  ($m, \omega_1, g_z$, etc.) not explicitly mentioned in Eq. (\ref{eq:exc}) are equal on both sides of the equation. 

Another general property of the Raman tensor in Eq.~(\ref{eq:R_lambda})  is that $R_{xx}$ is odd in $m$, while $R_{xy}$ is even in $m$.
This can be understood from symmetry arguments.
To begin with, the relation $h_{\rm e}(-{\bf k}, -m) = -h_{\rm e}({\bf k},m)$ implies that positive- and negative-energy bands are exchanged under the simultaneous operation ${\bf k}\to -{\bf k}$ and $m\to -m$. Replacing this result in Eqs.~(\ref{eq:ramanperturbation}) and (\ref{eq:diamagnetic}) of Appendix \ref{ap:main}, and using the fact that the electronic energies are invariant under ${\bf k}\to -{\bf k}$ (this is a consequence of the rotational symmetry of the model),  we get (for fixed $g_z$)
\begin{equation}
\label{eq:rel1}
R_{i j}(m,\eta) = -R_{ij}^*(-m,-\eta),
\end{equation}
which, in combination with Eq.~(\ref{eq:exc}), yields (for fixed $g_z$)
\begin{equation}
\label{eq:rel2}
R_{i j}(m,\eta) = -R_{ji}(-m,\eta).
\end{equation} 
From here, $R_{xx}(m,\eta) = -R_{xx}(-m,\eta)$ and $R_{xy}(m,\eta) = -R_{yx}(-m,\eta)= R_{xy}(-m,\eta)$, thereby proving the parities of $R_{xx}$ and $R_{xy}$ under the sign inversion of the Dirac mass.




Having discussed some generic properties of the Raman tensor, we next move to a more quantitative analysis. 
The explicit but cumbersome analytical expressions for $R_{xx}$ and $R_{xy}$ are shown in Appendix \ref{ap:ratio}.
As expected from gauge invariance \cite{basko2009calculation}, both $R_{xx}$ and $R_{xy}$ vanish when $\omega_1=\omega_0=0$.
Although $R_{xx}$ and $R_{xy}$ are rather complicated functions, their {\em ratio} becomes remarkably simple when $\omega_1\gg \omega_0$ ($\omega_1/\omega_0 \sim 10^2$ in typical Raman scattering experiments) and $\eta\to 0^+$:
\begin{equation}
\label{eq:ratio}
\frac{R_{xy}}{R_{xx}}= \frac{i \omega_1 v_y}{ 2 m v_x } + O\left(\frac{\omega_0}{\omega_1}\right) .
\end{equation}
This is the most salient result of our work.
It is unusual that such a simple functional relation should hold both when $\omega_1< 2 |m|$ and $\omega_1> 2|m|$.
For the purposes of verification, we have rederived Eq.~(\ref{eq:ratio}) in an alternative way, by recognizing that the Raman tensor describes the change in the electric polarization due to lattice vibrations.
Because the lattice vibrations that interest us here modulate the mass $m$, it follows (in the regime of negligible $\omega_0$) that
\begin{equation}
\label{eq:ratio_alter}
\frac{R_{xy}}{R_{xx}} = \frac{\partial \sigma_{xy}/\partial m}{\partial\sigma_{xx}/\partial m},
\end{equation}
where $\sigma_{ij}$ are elements of the frequency-dependent, complex optical conductivity.
The expressions for $\sigma_{i j}$ in our model of interest are available in the literature; see e.g. Ref. \cite{tse2010giant}.
Substituting those expressions in Eq.~(\ref{eq:ratio_alter}) and assuming that the ultraviolet cutoff energy  in Ref.~\cite{tse2010giant} is large compared to $\omega_1$ \footnote{For real materials it is not obvious that the ultraviolet cutoff energy (below which the electronic structure can be approximated by Eq.~(\ref{eq:He})) is large compared to the typical laser frequency; we will return briefly to this point later in the section.}, we recover Eq.~(\ref{eq:ratio}) for both $\omega_1>2|m|$ and $\omega_1<2 |m|$. Interestingly, the ratio $\sigma_{xy}/\sigma_{xx}$ is by no means simple.

The simplicity of Eq.~(\ref{eq:ratio})  has some interesting observable consequences. For example, when $\omega_1= 2 |m v_x/v_y|$, a selection rule for the Raman tensor emerges for circularly polarized light, i.e.
\begin{equation}
\label{eq:sel}
\hat{\bf e}_1^\dagger\cdot {\bf R}_\lambda\cdot\hat{\bf e}_2 = R_{xx} \left[1\mp {\rm sign}(v_x v_y m)\right]
\end{equation}
for $\hat{\bf e}_1 = \hat{\bf e}_2 = (\hat{\bf x} \pm i\hat{\bf y})/\sqrt{2}$
\footnote{If the incident and scattered lights have opposite circular polarizations, the Raman intensity vanishes altogether because out-of-plane phonons do not have any angular momentum. This is consistent with the helicity selection rule observed in transition metal dichalcogenides \cite{chen2015helicity}.}.
Hence, the Raman scattering amplitude vanishes when ${\rm sign}(v_x v_y m) = \pm 1$ (the top and bottom signs are for right- and left-circularly polarized photons, respectively).
This result, displayed in Fig.~\ref{fig:selection}, can be considered as a generalization to Raman scattering of the well-known valley selection rule in optical absorption \cite{wang2008valley,zeng2012valley,mak2018light}.
The latter takes place at resonance  (namely when the incident photon frequency $\omega_1$ coincides with the energy gap $2 |m|$ of the Dirac insulator), irrespective of the ratio $v_x/v_y$. In addition, if $v_x\neq v_y$, it can be readily shown that the optical valley selection rule applies to elliptically (rather than circularly) polarized photons.

The Raman selection rule emerging from Eq.~(\ref{eq:sel}) is different from the optical  valley selection rule in various ways. First, it happens at $\omega_1= 2 |m v_x/v_y|$, which is off resonance when $|v_x|\neq |v_y|$.  Second, 
it happens for circularly polarized photons, irrespective of the ratio $v_x/v_y$. Third, for $\eta\ll \omega_0$,  the resonance peak in the Raman intensity is split into two: for example, when $v_x=v_y$, one peak is at $\omega_1= 2 |m|$ and the other peak is at $\omega_1= 2|m|+\omega_0$ (only one peak is noticeable in Fig. \ref{fig:selection}, due to the energy smearing caused by $\eta$). There is no peak at $\omega_1= 2|m|-\omega_0$ because only Stokes processes are allowed at zero temperature. Lastly, while the conventional optical valley  selection rule is associated to  the real part of the optical conductivity (photon absorption process),
the selection rule in Eq. (\ref{eq:sel}) involves the sum of both the real and imaginary parts of the Raman tensor elements.
 
 
Another interesting consequence of Eq. (\ref{eq:ratio}) is that it implies a very simple relation between the phases of the Raman tensor elements:
\begin{align}
\label{eq:phase_diff}
 \phi_{xy}- \phi_{xx}&=\frac{\pi}{2} {\rm sign}(v_x v_y m),
\end{align}
where $\exp(i \phi_{ab}) \equiv R_{ab}/|R_{ab}|$.
While $\phi_{xx}$ and $\phi_{xy}$ are not observable on their own (for the simple reason that a {\em global} phase factor of the Raman tensor is not observable), their difference is observable as we explain below.
Remarkably, this phase difference is independent of the magnitudes of $\omega_1$, $v_x$, $v_y$ and $m$. 
In addition, the ``partial'' electronic Chern number of the 2D Dirac fermion model in the continuum, $ {\rm sign}(v_x v_y m)/2$, prominently intervenes in Eq.~(\ref{eq:phase_diff}).

\begin{figure}[t]
\begin{center}
   \includegraphics[width=1\columnwidth]{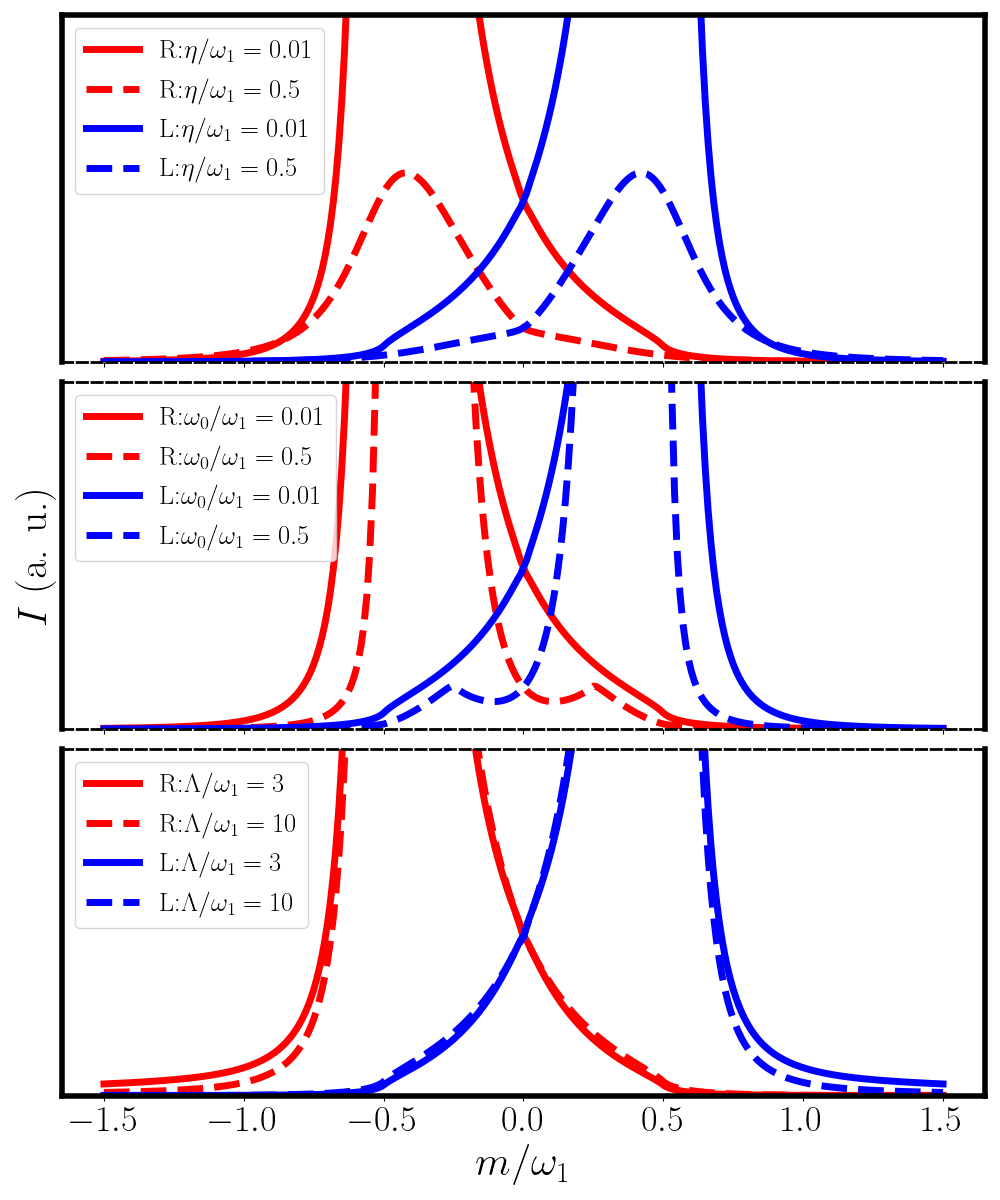}                            
\caption{Effect of varying the electronic decay rate $\eta$ (top panel), the phonon frequency $\omega_0$  (middle panel) and the ultraviolet cutoff $\Lambda$ (bottom panel) in the results of Fig.~\ref{fig:selection}, for isotropic Fermi velocities ($v_x=v_y$). Unless indicated otherwise, $\omega_0 =0.01 \omega_1$, $\eta=0.01 \omega_1$ and $\Lambda=\infty$.} 
 \label{fig:selection2}
  \end{center}
\end{figure} 
In order to emphasize that the simplicity of Eq.~(\ref{eq:phase_diff}) is nontrivial, Fig. \ref{fig:rxxrxy} illustrates $\phi_{xx}$ and $\phi_{xy}$ separately: each has a relatively complicated dependence on $m/\omega_1$, but their difference is a step function.
When $\omega_1< 2|m|$  (i.e. when there is no real photon absorption), the behavior of $\phi_{xx}$ and $\phi_{xy}$ can be readily understood. 
In this regime, the infinitesimal $\eta$ term can be altogether ignored. Then, from setting $\eta=0$ in Eq.~(\ref{eq:exc}), we get that $R_{xx}$ is a real number and $\phi_{xx}$ equals either $0$ or $\pi$ (mod $2\pi$). In addition, as mentioned above, $R_{xx}$ is odd in $m$,
which implies that $\phi_{xx}$ will shift by $\pi$ when the sign of $m$ is switched. 
In contrast, $R_{xy}$ is purely imaginary when $\omega_1< 2|m|$; this is again a consequence of Eq.~(\ref{eq:exc}) and the rotational symmetry. 
Hence, $\phi_{xy}$ equals either $\pi/2$ or $-\pi/2$ (mod $2\pi$).
In addition, because $R_{xy}$ is even in $m$, $\phi_{xy}$ remains unchanged when the sign of $m$ is switched. 
In sum, when $\omega_1<2|m|$, $\phi_{xy}-\phi_{xx}$ must equal $\pi/2$ or $-\pi/2$, with a sign reversal under $m\to -m$.

When $\omega_1>2 |m|$, the behavior of $\phi_{xx}$ and $\phi_{xy}$ becomes more complicated but, surprisingly, $\phi_{xx}-\phi_{xy}$ is still quantized to $\pm \pi/2$.  In that regime, any infinitesimal electronic decay rate $\eta$ leads to a sizable imaginary part for $R_{xx}$ and a sizable real part for $R_{xy}$, both of which vary as a function of $\omega_1$. Accordingly, $\phi_{xx}$ and $\phi_{xy}$ vary with $m$. However, as per Eq.~(\ref{eq:rel2}),  $\phi_{xx}$ still shifts by $\pi$ under $m\to -m$, while $\phi_{xy}$ remains invariant; this explains why $\phi_{xx}-\phi_{x y}$ jumps by $\pi$ at  $m=0$.
In contrast, we have no simple argument for why $\phi_{xx}-\phi_{xy}$ remains quantized when $\omega_1>2 |m|$.

\begin{figure}[t]
\begin{center}
   \includegraphics[width=\columnwidth]{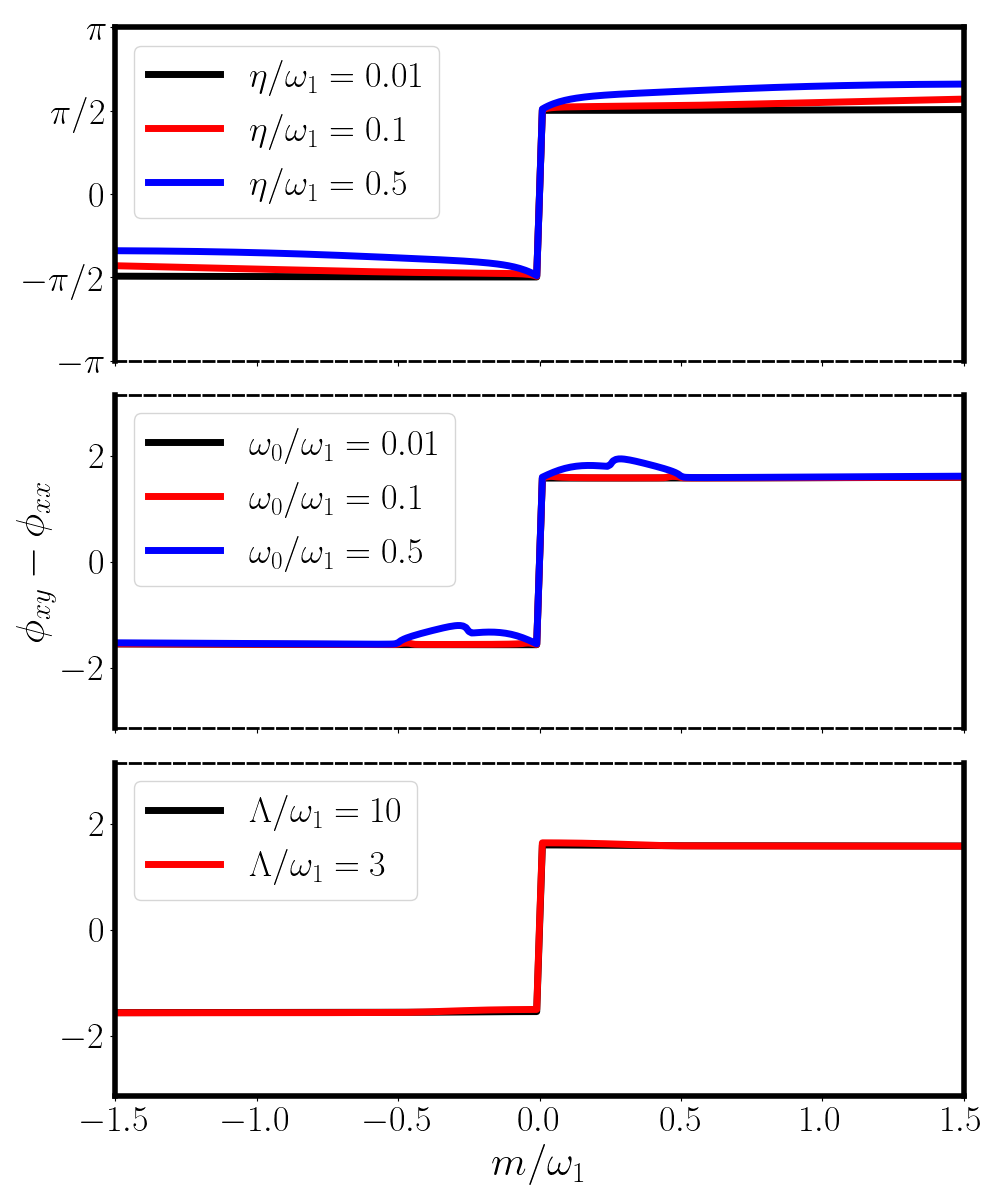}                            
\caption{Effect of varying the electronic decay rate $\eta$ (top panel), the phonon frequency $\omega_0$  (middle panel) and the ultraviolet cutoff $\Lambda$ (bottom panel) in the results of Fig.~\ref{fig:rxxrxy}.
Unless indicated otherwise, $\omega_0 =0.01 \omega_1$, $\eta=0.01 \omega_1$ and $\Lambda=\infty$.}
 \label{fig:phasedif2}
  \end{center}
\end{figure}

Because Eq.~(\ref{eq:ratio}) has been obtained for an infinitesimal electronic decay rate $\eta$, a negligible phonon frequency $\omega_0$ and an infinite UV cutoff $\Lambda$, it is necessary to ask how other values of $\eta$, $\omega_0$ and $\Lambda$  would modify the result. 
Figures \ref{fig:selection2} and \ref{fig:phasedif2} confirm that the Raman selection rule and the simple step like behavior in  $\phi_{xx}- \phi_{xy}$ as a function of $m$ are robust when $\eta$ and $\omega_0$ are given a range of reasonable values.
In addition, the results are not strongly sensitive to the value of $\Lambda$, insofar as $\Lambda\gg \omega_1$ (the starting electronic model would be of dubious relevance if we took $\Lambda\lesssim \omega_1$).
In particular, the discontinuity of $\phi_{xx}-\phi_{x y}$ by $\pi$ at $m=0$ holds irrespective of the values of $\eta$, $\omega_0$, and $\Lambda$ (see Fig. \ref{fig:phasedif2}), due to the general relation (\ref{eq:rel2}). 

\begin{figure}[t]
\begin{center}
   \includegraphics[width=1\columnwidth]{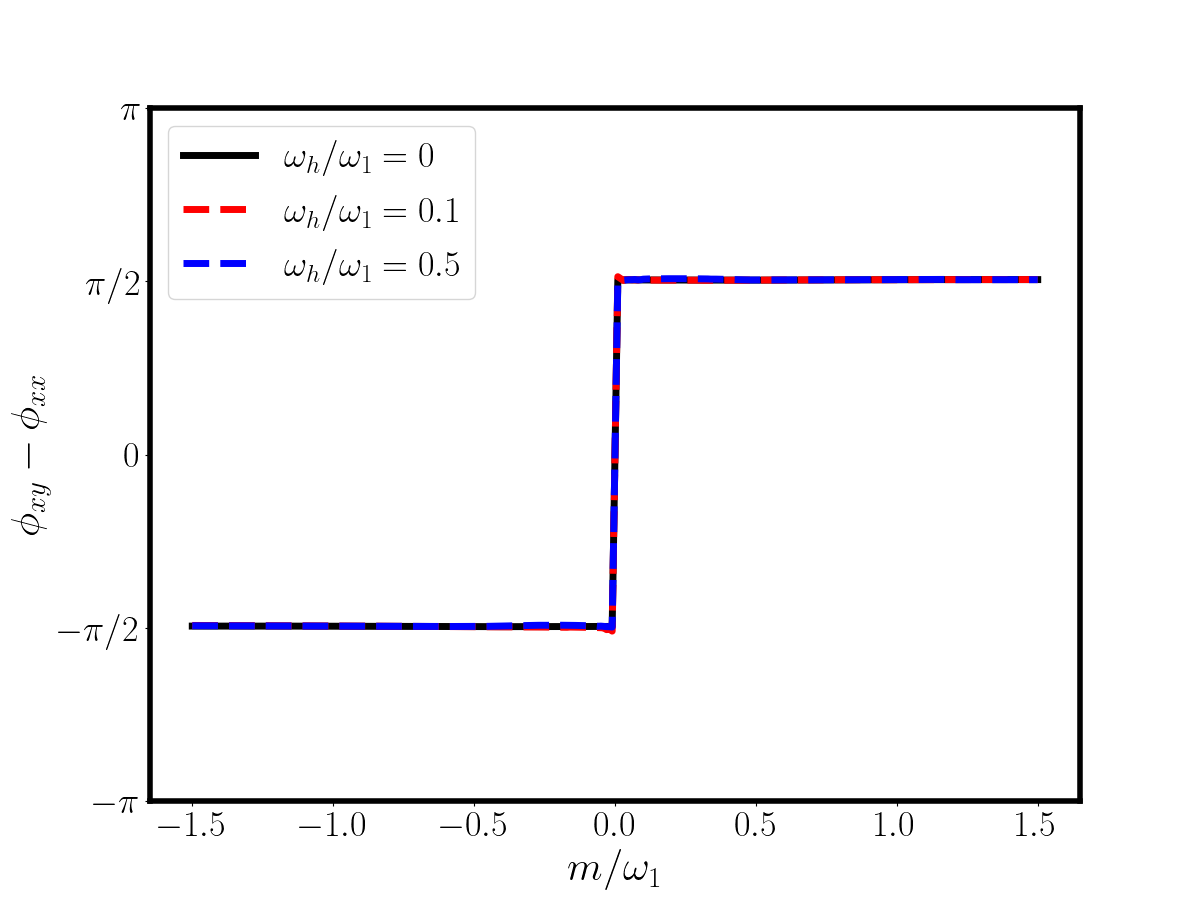}                            
\caption{Effect of hexagonal warping (of amplitude $\omega_h$) on the results of Fig.~\ref{fig:rxxrxy}. The curves are obtained for $\Lambda=50 \omega_1$, $\omega_0 =0.01 \omega_1$, $\eta=0.01 \omega_1$. Unlike in Fig.~\ref{fig:rxxrxy}, we find that a finite cutoff is needed to avoid numerical artifacts when $\omega_h\neq 0$; the results are nevertheless independent of the value of $\Lambda$. } 
 \label{fig:kcubewarping}
  \end{center}
\end{figure} 



Aside from $\eta$, $\omega_0$ and $\Lambda$, our results could arguably be modified by (in practice, inevitable) perturbations that break the full rotational symmetry of the model. To test this, we have added to $h_{\rm e}({\bf k}, m)$ a hexagonal warping term $\omega_h (k_+^3+k_-^3) \sigma_z$, where $k_\pm=k_x\pm i k_y$ and $\omega_h$ is a real constant \cite{fu2009hexagonal}. This term, known to arise on the surface states of three-dimensional topological insulators, reduces the rotational symmetry of the model to $C_3$ but does not affect the Chern number of the massive Dirac bands. 
In Fig.~\ref{fig:kcubewarping}, we plot $\phi_{xx}-\phi_{x y}$ for different values of $\omega_h$. 
Clearly, the main features of $\phi_{xx}-\phi_{x y}$ are robust to hexagonal warping. 

We have likewise considered perturbations of the type $D k^2 \sigma_0$ and $(t_x k_x +t_y k_y)\sigma_0$, proportional to the identity matrix,  in $h_{\rm e}({\bf k}, m)$. Here, $D$, $t_x$, and $t_y$ are real numbers. While these terms break particle-hole symmetry and create a tilted energy dispersion, they have no effect in the Berry curvature and hence on the Chern number. Furthermore, it can be readily shown from the microscopic theory  that terms proportional to $\sigma_0$ have no impact in the Raman tensor insofar as we maintain the Fermi energy inside the energy gap and insofar as the temperature is zero (cf. Appendix~\ref{ap:main}). 

For comparison, we have also calculated the Raman tensor for the geometrically trivial electronic models $h_{\rm e}({\bf k}, m)=(m+k^2) \sigma_z$ and $h_{\rm e}({\bf k}, m)=k^2 \sigma_z+m \sigma_x$.
The electronic bands in these models have vanishing Berry curvature and Chern number because only two Pauli matrices appear in the Hamiltonians. It is easy to see that these models result in $R_{xy}=0$, thereby preventing the two physical consequences of Eq.~(\ref{eq:ratio}). Had we considered  $h_{\rm e}({\bf k}, m)=k_x \sigma_x+k_y\sigma_y$, we would find $R_{xx}=0$ instead, again preventing the two physical consequences of Eq.~(\ref{eq:ratio}). Therefore, for the special selection rule and phase difference to take place, it appears that two-band electronic models in which all three Pauli matrices intervene is a minimal essential ingredient. This condition, together with rotational symmetry in the $xy$ plane, makes the massive Dirac Hamiltonian the minimal model for which the aforementioned interesting features of the Raman tensor are realized.

In view of the preceding analysis, the (half-) quantization of $(\phi_{xy} -\phi_{xx})/\pi$ across the entire spectrum of $\omega_1$ could suggest an underlying topological argument to explain it.
Yet, we have not succeeded in finding such an argument.
Instead, to decide whether the appearance of the Chern number in Eq.~(\ref{eq:phase_diff}) is fundamental or fortuitous, we have studied a variant of the electronic model by adding a   ``Newtonian" mass term $B k^2 \sigma_z$ to $h_{\rm e}({\bf k}, m)$.
In that model, the Chern number of the occupied band is 
\begin{equation}
C={\rm sign}(v_x v_y) \left[{\rm sign}(m) -{\rm sign}(B)\right]/2.
\end{equation}  


When $B\neq 0$, several results discussed thus far are modified.
For instance, Eqs.~(\ref{eq:rel1}) and (\ref{eq:rel2}) are straightforwardly generalized to
\begin{align}
\label{eq:rel3}
R_{i j}(m, B, \eta) &= -R_{ij}^*(-m, -B, -\eta)\nonumber\\
R_{i j}(m, B, \eta) &= -R_{ji}(-m, -B, \eta).
\end{align}
In addition, when $B\neq 0$, a ``diamagnetic contribution'' to the Raman tensor emerges, which cannot be neglected (see Appendix \ref{ap:main}). 
Finally,  Eq.~(\ref{eq:ratio}) is no longer correct when $B\neq 0$; the corresponding expression for $B\neq 0$ is no longer simple and we will not write it here (but see Appendix \ref{ap:finiteB} for partial analytical results).  

\begin{figure}[t]
\begin{center}
   \includegraphics[width=\columnwidth]{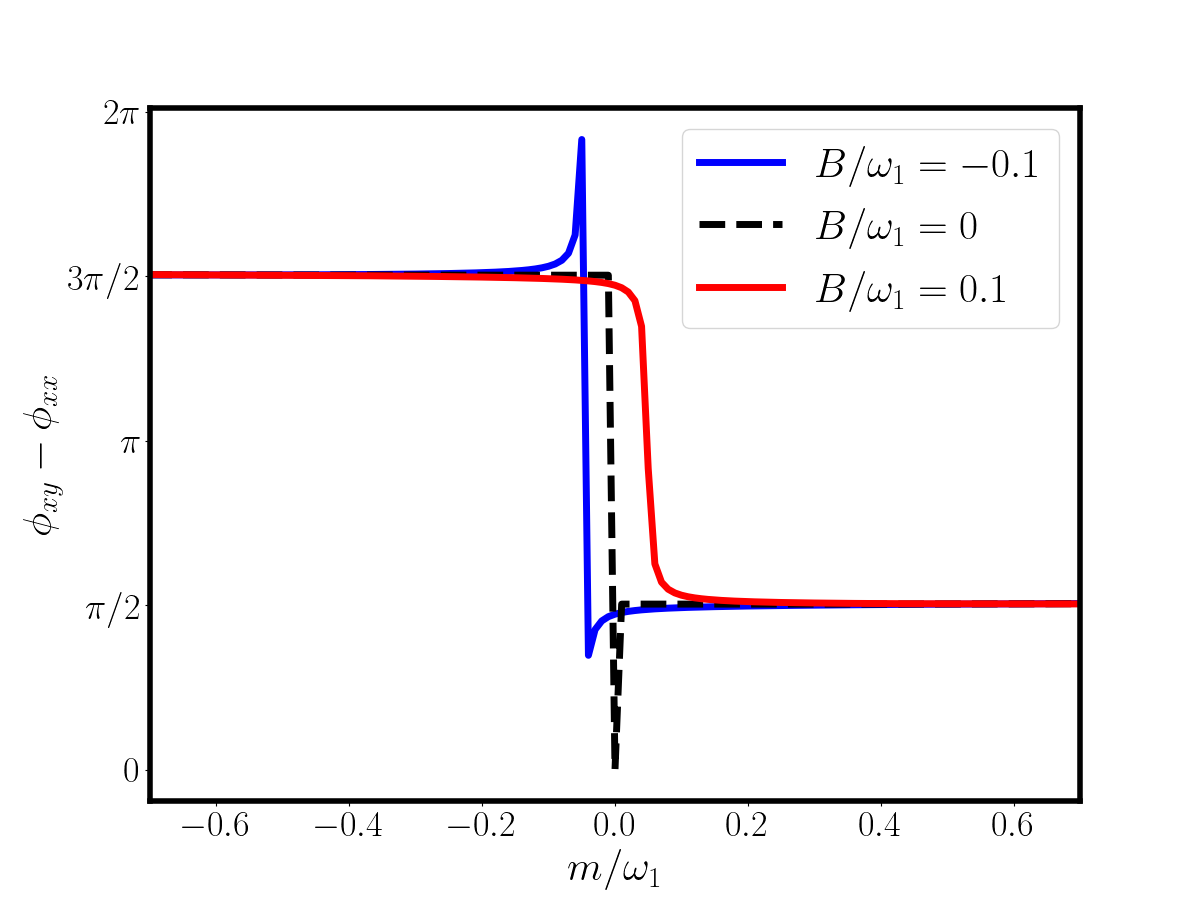}                            
\caption{Influence of a Newtonian mass term (added to the electronic Hamiltonian as $B k^2 \sigma_z$) in the phase difference between the $R_{xx}$ and $R_{xy}$ Raman tensor elements, as a function of  the ratio between the Dirac mass $m$ and the incident photon frequency $\omega_1$. 
The phonon energy and the electronic decay rate are set to $\omega_0 =0.01 \omega_1$ and $\eta=0.01 \omega_1$, respectively. 
}
 \label{fig:Bterm}
  \end{center}
\end{figure} 

In spite of the aforementioned changes, the step structure of $\phi_{xy} -\phi_{xx}$ as a function of $m/\omega_1$ remains strikingly robust (Fig. \ref{fig:Bterm}). 
In particular, if $B$ and $m$ are taken to be proportional to one another, the step is located at $m=B=0$.
This is a direct consequence of the second line of Eq.~(\ref{eq:rel3}).
If only $m$ is varied while  $B$ is kept fixed, the step in $\phi_{xy} -\phi_{xx}$ is shifted to  $m=\gamma B$, where $\gamma$ is a  positive number. 

The preceding findings indicate that $(\phi_{xy} -\phi_{xx})/\pi$ cannot be identified with the electronic Chern number $C$, as otherwise it would be equal to $0$ or $1$ (instead of $-1/2$ or $1/2$).
In addition, the hypothesis that the jump in $(\phi_{xy} -\phi_{xx})/\pi$ could correspond to a change of $C$ is refuted by 
the fact that the Chern number changes discontinuously across $m=0$ irrespective of the value of $B$, whereas the discontinuity in $(\phi_{xy} -\phi_{xx})/\pi$  is located at $m=0$ only when $B=0$.
In Appendix \ref{ap:finiteB}, we show analytical results for $R_{xx}$ and $R_{xy}$ in the regime $\omega_1\ll 2 |m|$, which show that Eq.~(\ref{eq:phase_diff}) applies in that regime regardless of the value of $B$. 
Again, this result disproves a direct relation between $C$ and $(\phi_{xy} -\phi_{xx})/\pi$.




The nonetheless intriguing step structure of  $\phi_{xy}-\phi_{xx}$ is potentially measurable through angle-resolved Raman spectroscopy \cite{strach1998determination,pimenta2021polarized}. 
First, for a fixed $\omega_1$, $|R_{xx}|$ and $|R_{xy}|$ can be extracted from the Raman intensities measured in the configurations  ($\hat{\bf e}_1, \hat{\bf e}_2) = (\hat{\bf x}, \hat{\bf x})$ and ($\hat{\bf e}_1, \hat{\bf e}_2) = (\hat{\bf x}, \hat{\bf y})$, respectively. Then, choosing $\hat {\bf e}_1=\hat{\bf e}_2=\sin \theta\hat{\bf x}+ i\cos \theta \hat{\bf y}$,  we have
 \begin{align}
 \label{eq:angle}
I_\lambda &\propto |R_{xx}|^2+|R_{xy}|^ 2\sin ^2(2\theta) \nonumber \\
&+2|R_{xx}| |R_{xy}| \sin(2\theta) \sin(\phi_{xx}-\phi_{xy}).
 \end{align}
Thus,  knowing $|R_{xx}|$ and $|R_{xy}|$ and changing the value of $\theta$, one can measure $\sin(\phi_{xx}-\phi_{xy})$.
Such measurement could be realized by carrying out Raman spectroscopy on a surface 
of a magnetic topological insulator \footnote{ Our theory applies to the bulk 2D electronic contribution to the Raman scattering, while disregarding the (relatively unimportant) contributions from the one-dimensional chiral edge states of the Chern insulator.}. 
Single Dirac fermions emerge therein, with a mass that switches sign under reversal of the magnetic order \cite{tokura2019magnetic}.
Said reversal should manifest in a sign change of the last term of Eq.~(\ref{eq:angle})
\footnote{This is true irrespective of whether or not $g_z$ switches sign when the magnetic order is reversed, because $\phi_{xy}-\phi_{xx}$ is independent of $g_z$.}.

In practice, one limitation of our theory is that it captures only the contribution from Dirac-like electrons to the Raman tensor. 
In a real material, higher-energy electronic states (which are not Dirac-like) also contribute to the Raman intensity, and those contributions can affect $\phi_{xy}-\phi_{xx}$ and possibly mask the simplicity of Eq.~(\ref{eq:phase_diff}). 

One possible solution to the preceding issue is to use resonant Raman scattering, where the incident photon frequency $\omega_1$ is close to the energy gap $2 |m|$ of the Dirac insulator. 
Then, the main contribution to the Raman tensor should emerge from the Dirac fermion states, thereby rendering our theory more relevant to experiment.
Unfortunately, the energy gaps of surface states in magnetized topological materials are of the order of a few meV \cite{tokura2019magnetic}, while $\omega_1\sim 1 {\rm eV}$ in conventional Raman scattering experiments.
Significantly reducing $\omega_1$ does not appear to be a viable solution, as it would entail a strong decrease of the overall Raman intensity 
\cite{cardona2005resonance}.

The aforementioned problems motivate us to extend our calculations to two-valley Dirac insulators, where our theory may have a closer application to experiment.

\begin{figure}[b]
  \begin{center}
    \includegraphics[width=\columnwidth]{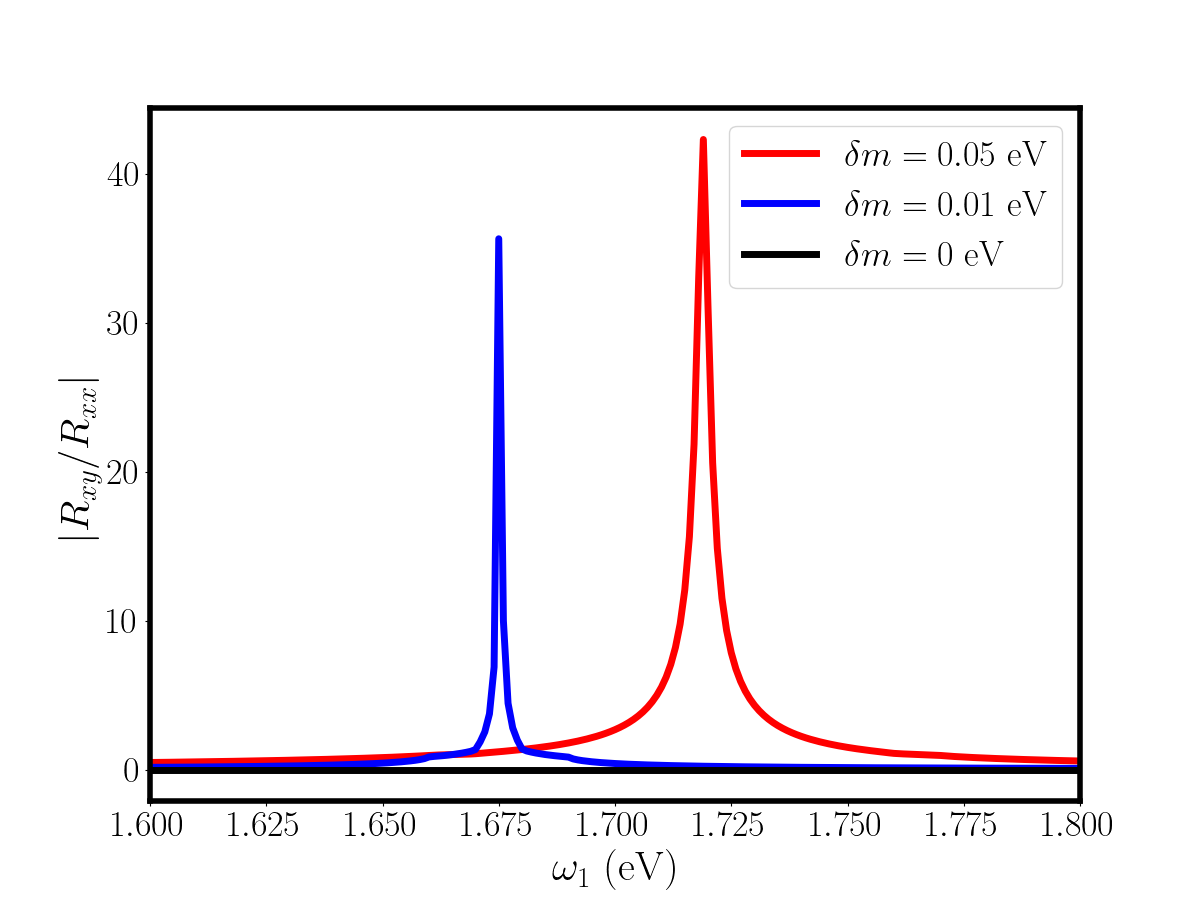}                            
\caption{Ratio between the moduli of the Raman tensor elements, computed for a low-energy model of two massive Dirac fermions. Broken time-reversal symmetry ($\delta m\neq 0$) is necessary in order to have a nonzero $|R_{xy}|$. 
The electronic  band parameters (except for $\delta m$) are taken from Ref.~\cite{xiao2012coupled}.  The electronic decay rate and the phonon frequency are chosen as $\eta=0.02 |\delta m|$ and $\omega_0 = 0.01 {\rm eV}$, respectively.}
 \label{fig:twovalleyratio}
  \end{center}
\end{figure}

 \section{Generalizations}
 \label{sec:generalizations}
 
 \subsection{Two Dirac fermions}
 \label{sec:two}

The simplest  low-energy electronic model for electrically insulating monolayers of transition metal dichalcogenides (TMDCs) such as MoS$_2$ consists of a pair of two-dimensional Dirac Hamiltonians \cite{xiao2012coupled}.
The Dirac fermions in TMDC monolayers are located in two separate valleys of the Brillouin zone (denoted $K$ and $K'$), which are related to one another by time reversal.
Since the energy gaps of these materials can be comparable to the typical values of $\omega_1$, it is feasible to use resonant Raman scattering to preferentially probe the contribution from Dirac fermions. 

To obtain the Raman tensor, we first calculate the partial Raman tensor for each valley as in Sec. \ref{sec:onevalley}, and then sum the results for the two valleys 
\footnote{ In the microscopic theory of the Raman tensor elements (App. A),
there is a sum over the electronic wave vector. In a crystal, this sum runs over the entire Brillouin zone. In the low-energy two-valley model with a UV cutoﬀ used in Sec. \ref{sec:two}, the sum over the wave vectors is truncated to the vicinities of the valleys. Then, the total Raman tensor can be decomposed into the sum of Raman contributions coming from
each valley. Each contribution matches that of a single Dirac fermion with a UV cutoff, which we have already studied in
Sec. \ref{sec:onevalley}. This is of course an approximated calculation, where the contribution from the high-energy (non Dirac-like) electronic states
to the Raman tensor is ignored. The approximation appears justifiable when the energies of the incoming and outgoing photons are close to the electronic bandgap.}. 
Like in Sec. \ref{sec:onevalley}, we adopt a fully rotationally symmetric electronic structure model (e.g., neglecting the hexagonal warping) in the $xy$ plane. Then, only out-of-plane phonons that modulate the Dirac mass contribute to the Raman tensor. 

Due to time-reversal symmetry, the Dirac  mass $m$ and the electron-phonon coupling $g_z$ take opposite values on the two valleys \footnote{The convention one finds often in the literature is that the Dirac masses in the two valleys are taken to be identical, while one component of the Dirac velocity (say, $v_x$) changes sign from one valley to another. Here, we take a different convention of keeping $v_x=v_y$ to be identical for the two valleys, yet reversing the sign of the Dirac mass between the valleys. The end results for the Raman tensor are the same in the two conventions.}.
As a result, the two valleys make perfectly canceling contributions to $R_{xy}$. This, in turn, implies that $\phi_{xy}-\phi_{xx}$ is not observable: for instance, the term of Eq.~(\ref{eq:angle}) containing the information about the phase difference vanishes for $R_{xy}=0$. 
To circumvent this problem, we consider a TMDC monolayer with broken time-reversal symmetry, which can be achieved, e.g., through proximity coupling to a magnetic insulator with an out-of-plane magnetization \footnote{A similar symmetry-breaking can be attained by an external magnetic field, but its influence in the band structure is believed to be smaller. In addition, the treatment of an external static magnetic field would require adding the effect of a vector potential in ${\cal H}_e$, thereby leading to Landau levels. We will not consider that situation in the present work.}.
The exchange coupling between the TMDC and the magnet 
breaks ${\cal T}$, causing a valley splitting  of $\sim 10\, {\rm meV}$ \cite{qi2015giant, lin2020magnetic, li2022giant, zollner2023strong, ahammed2022valley}  that has been observed in experiment \cite{ciorciaro2020observation, zhao2017enhanced, zhong2017van, zhang2020controllable, norden2019giant, ge2022enhanced, choi2023asymmetric}.
In our calculation, we model the valley splitting by adopting a mass $m$ for  valley $K$  and a mass $-m -\delta m$ for the valley $K'$, with $|\delta m|\ll |m|$.
Consequently, the two valleys no longer make perfectly canceling contributions to $R_{xy}$ (see Fig.~\ref{fig:twovalleyratio}).
We neglect the difference in $|g_z|$ between the two valleys.

\begin{figure}[t]
  \begin{center}
    \includegraphics[width=\columnwidth]{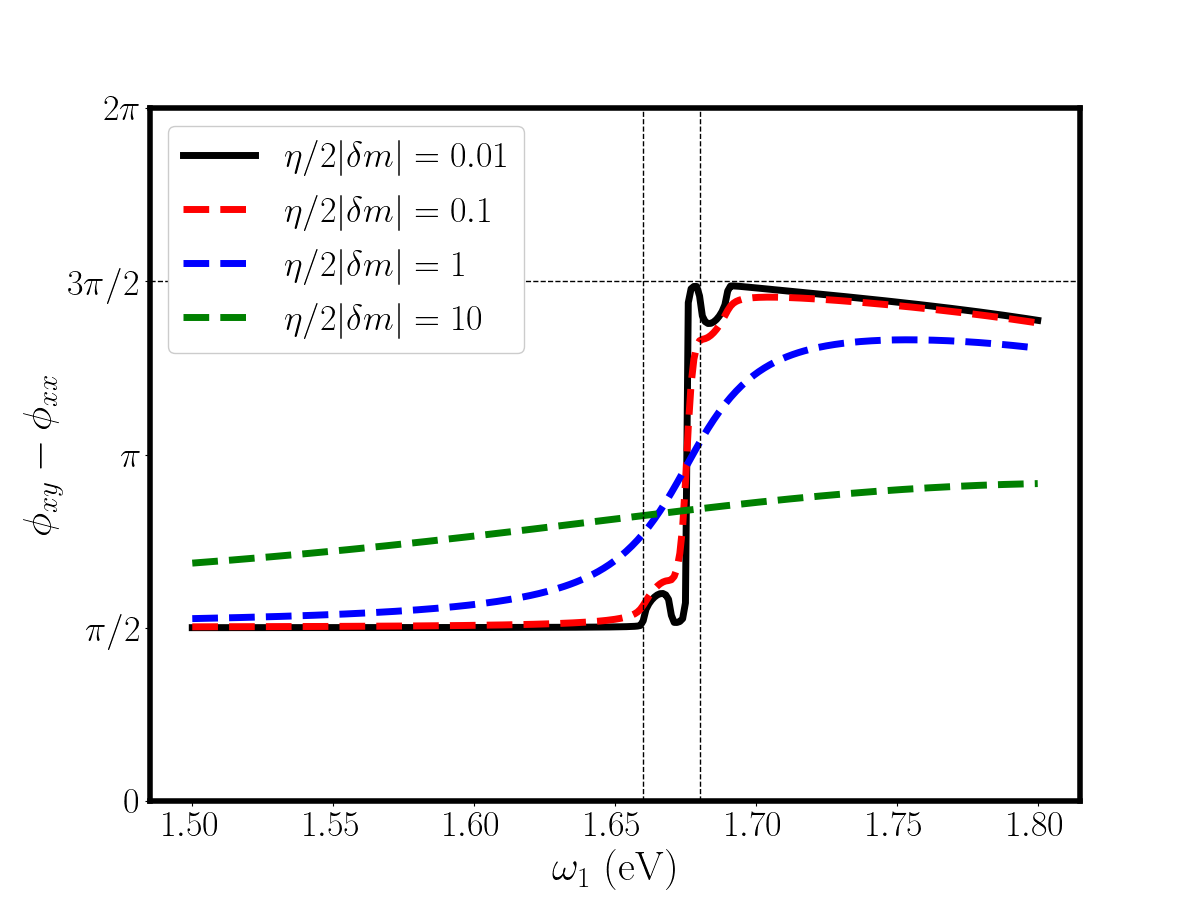}                            
\caption{Phase difference between the $R_{xx}$ and $R_{xy}$ Raman tensor elements as a function of the incident light frequency, in a low-energy model of MoS$_2$
 with an out-of plane proximitized magnetism. The electronic  band parameters (except for the time-reversal-breaking term $\delta m=0.01 {\rm eV}$) are taken from Ref.~\cite{xiao2012coupled}.
The vertical dashed lines represent the two resonant frequencies ($\omega_1 = 2 |m|$ and $\omega_1=2 |-m-\delta m|$), in which the incident light frequency matches with the energy gap at one of the valleys.  The phonon energy is $\omega_0 =0.01 {\rm eV}$.  
When the electronic decay rate $\eta$ is small, the value of the phase difference at the two resonances matches with the expectation from Eq.~(\ref{eq:phase_diff}). 
The small notches near resonance are a product of $\omega_0$. 
When $\eta$ becomes comparable to $|\delta m|$, the quantization of the phase difference near resonance is lost.}
 \label{fig:twovalleyphase}
  \end{center}
\end{figure}

When $\delta m\neq 0$, the $A_1'$ phonon of MoS$_2$ (which corresponds to out-of-plane lattice vibrations) admits an antisymmetric Raman tensor.
To see this, we begin by noting that the normal coordinate associated to the $A_1'$ mode remains invariant under all symmetry operations of the crystal's point group. 
Therefore, to find out the symmetry-imposed constraints on the Raman tensor elements, we can use Eq.~(\ref{eq:neumann1}) with $\lambda=A_1'$. 
The $C_3$ axis perpendicular to the 2D plane enforces $R_{xx} = R_{yy}$ and $R_{xy} = -R_{yx}$.
 This follows from replacing $O_{xx}=O_{yy}=-1/2$ and $O_{xy}=-O_{yx}=-\sqrt{3}/2$ in Eq.~(\ref{eq:neumann1}).
In the absence of a magnetic perturbation ($\delta m=0$), the vertical mirror planes of the $D_{3h}$ point group would further impose $R_{xy}=0$.
This can be seen, e.g., by replacing $O_{xx}=-O_{yy}=-1$ and $O_{xy}=O_{yx}=0$ in Eq.~(\ref{eq:neumann1}).
Yet, an out-of-plane magnetic perturbation ($\delta m\neq 0$) breaks all vertical mirror planes, as well as the horizontal $C_2$ axes. 
As a result, only the $C_3$ axis along $z$ and the horizontal mirror plane remain as symmetry operations (point group $C_{3h}$). 
The horizontal mirror plane acts trivially (as an identity matrix) in Eq.~(\ref{eq:neumann1}); as such, it does not create any constraints on $R_{ij}$ and we end up with an antisymmetric Raman tensor dictated by the $C_3$ axis ($R_{xx}= R_{yy}$ and $R_{xy} = -R_{yx}$).
One can likewise show that, the $A_2'$ phonon mode, which is Raman silent in the $D_{3h}$ point group, becomes Raman active in the $C_{3h}$ point group, with an antisymmetric Raman tensor.
 
 \begin{figure}[t]
  \begin{center}
    \includegraphics[width=\columnwidth]{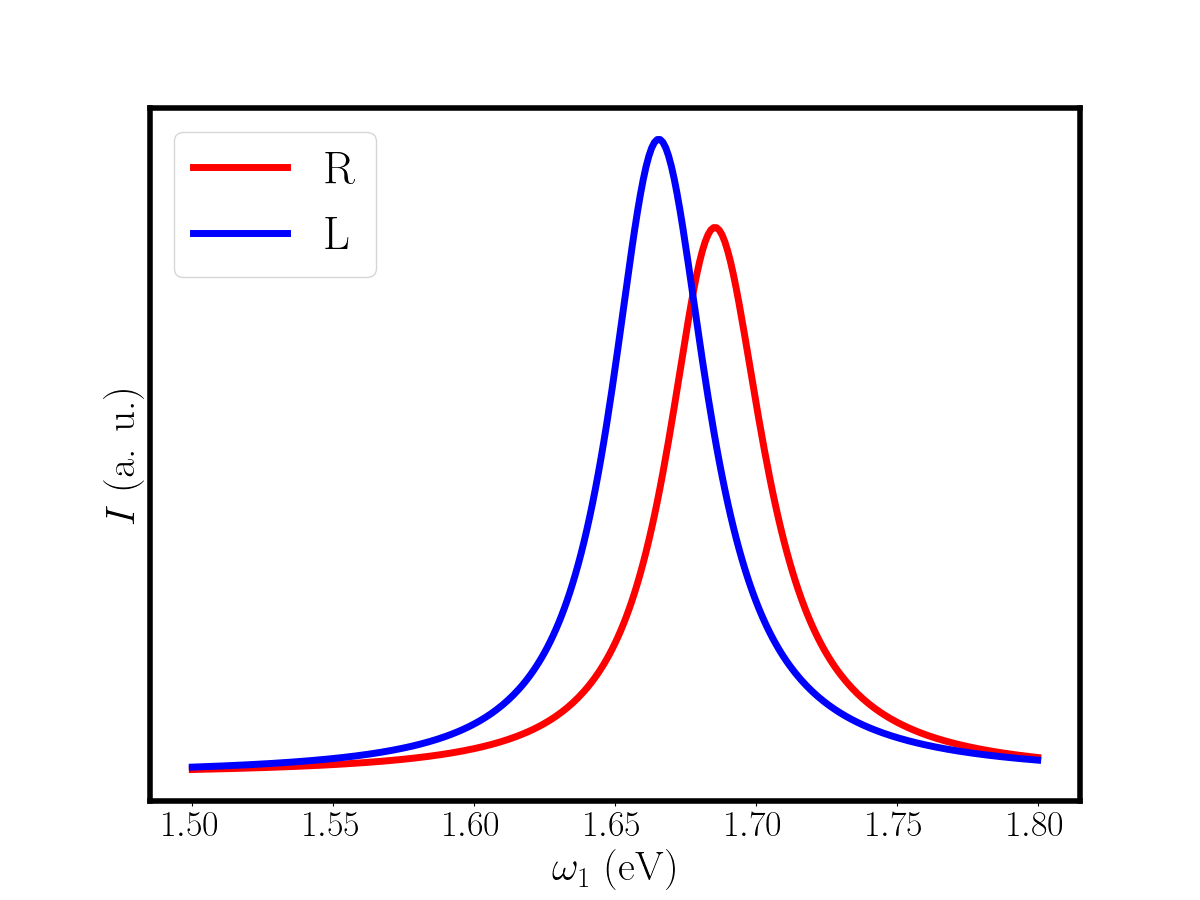}                            
\caption{Raman scattering intensity off an out-of-plane phonon as a function of the incident light frequency, in a low-energy model of MoS$_2$
 with an out-of plane proximitized magnetism. The red and blue curves correspond to right- and left-circularly polarized light, respectively, with the same polarization for incident and scattered light. The difference in the two curves, which originates from broken time-reversal symmetry ($\delta m = 0.01 {\rm eV}$), is noticeable even when the electronic decay rate is significant ($\eta=2 |\delta m|$ in the figure). The electronic  band parameters (except for $\delta m$) are taken from Ref.~\cite{xiao2012coupled}.}
 \label{fig:twovalleyselection}
  \end{center}
\end{figure}


\begin{figure*}[t]
  \begin{center}
    \includegraphics[width=\textwidth]{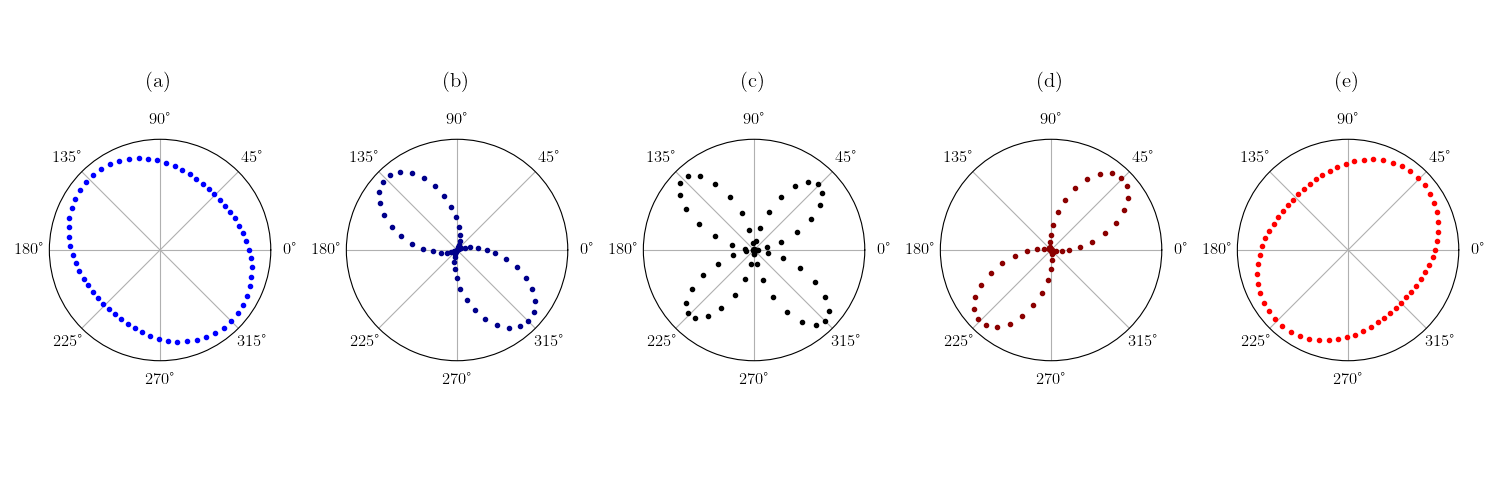}                            
\caption{Angular $(\theta$) dependence of the Raman scattering intensity in a low-energy model of MoS$_2$
 with an out-of plane proximitized magnetism, when the incident and scattered lights are polarized as $\hat {\bf e}_1=\hat{\bf e}_2=\sin \theta\hat{\bf x}+ i\cos \theta \hat{\bf y}$; see main text for description.
(a) $\omega_1=1.5\, {\rm eV}$,  (b) $\omega_1=1.66\,  {\rm eV}$ (resonance at valley $K$),  (c) $\omega_1=1.675\,  {\rm eV}$,  (d) $\omega_1=1.68\,  {\rm eV}$ (resonance at valley $K'$) , (e) $\omega_1=1.8\,  {\rm eV}$. The parameter values adopted are those of Fig.~\ref{fig:twovalleyphase}, with $\eta=0.02 |\delta m|$. 
The intensity plots for different values of $\omega_1$ are not to scale.} 
 \label{fig:angle_resolved}
  \end{center}
\end{figure*}

The simple ratio of Eq.~(\ref{eq:ratio}) holds for the partial Raman tensor of each valley, but not for the total Raman tensor summed over the valleys. For the latter, we find
\begin{equation}
\frac{R_{xy}}{R_{xx}}\simeq\frac{i \omega_1}{2m} \frac{\delta m}{2} \frac{\partial \ln (R_{xy,K})}{\partial m}
\end{equation}
to first order in $\delta m/m$. Here, $R_{xy,K}$ is the contribution from valley $K$ to $R_{xy}$.

The fact that Eq.~(\ref{eq:ratio}) no longer holds for the total Raman tensor elements of a two-valley Dirac system might induce one to think that the phase difference $\phi_{xy}-
\phi_{xx}$ ought not have any simple features. That is not the case, however, as we now discuss.

Figure \ref{fig:twovalleyphase} displays $\phi_{xy}-\phi_{xx}$ as a function of $\omega_1$ for a low-energy model of a magnetized monolayer of MoS$_2$, with $B=0$ (no Newtonian mass) and $\delta m\neq 0$. 
Because the energy gaps in the two valleys are different, there are two distinct resonance frequencies, i.e. $\omega_1 = 2 |m|$ and $\omega_1=2|-m-\delta m|$.
When $\omega_1$ is below the smallest energy gap and $\eta$ is small, $\phi_{xy}-\phi_{xx}$ is quantized to $\pi/2$ because $R_{xx}$ is purely real and $R_{xy}$ is purely imaginary (the reason for this has been explained in the preceding section).

Soon after $\omega_1$ exceeds the lowest resonance frequency,  $\phi_{xy}-\phi_{xx}$ abruptly changes and reaches $-\pi/2$ (or equivalently $3\pi/2$) at the second resonance. 
This change of sign  in $\phi_{xy}-\phi_{xx}$ is consistent with Eq.~(\ref{eq:phase_diff}), as the two valleys have opposite signs of the Dirac mass. 
It is as though, close to the first (second) resonance, the main contribution to the Raman tensor came from the first (second) valley.
That is, $R_{ij} \simeq R_{i j, K}$ when $\omega_1 \simeq 2|m|$ and $R_{ij} \simeq R_{i j, K'}$ when $\omega_1 \simeq 2|-m-\delta m|$, which is a sensible result.
Since Eq.~(\ref{eq:ratio}) applies to $R_{i j, K}$ and $R_{i j, K'}$ separately, it follows that the total Raman tensor will also follow Eqs. (\ref{eq:ratio}) and (\ref{eq:phase_diff}), albeit only in the vicinity of a resonance.
Hence, $\phi_{xy}-\phi_{xx}$ remains a probe of the sign of the Dirac mass in TMDC monolayers (as per the convention described in the preceding footnote).
This is the main first result from the present section.

Beyond the second resonance (i.e. when $\omega_1$ exceeds the largest energy gap), $\phi_{xy}-\phi_{xx}$ continuously changes with $\omega_1$. In this regime, the two valleys contribute comparably to the Raman tensor, and thus Eq.~(\ref{eq:phase_diff}) is not expected to apply.

As the value of $\eta$ is increased in Fig. \ref{fig:twovalleyphase}, the dependence of $\phi_{xy}-\phi_{xx}$ on $\omega_1$ gets smoother, the step structure being largely washed out by the time $\eta\gg |\delta m|$.
In experiment, $\eta\simeq |\delta m|$ has been attained (see e.g. Ref. \cite{ciorciaro2020observation}).

To test the robustness of our results, we have added Newtonian mass and trigonal warping terms that emerge in the density functional theory (DFT)-informed low-energy effective Hamiltonian of monolayer MoS$_2$  \cite{kormanyos2013monolayer}.
We find that the results of  Fig. \ref{fig:twovalleyphase} still hold in this extended model, provided that a UV cutoff $\Lambda^*$ is introduced to ensure that the $O(k^2)$ and $O(k^3$) terms in the electronic structure remain small compared to the $O(k)$ term.  Unlike in Sec.~\ref{sec:onevalley}, our results for the Raman tensor obtained from the model of Ref.  \cite{kormanyos2013monolayer}  depend on the value of the cutoff if the latter exceeds $\Lambda^*$. 
This implies the need to resort to lattice models in order to obtain more reliable results. While this task is beyond the scope of the present work, our preliminary results indicate that the curves in Fig. \ref{fig:twovalleyphase} are reproduced in a tight-binding model on a honeycomb lattice with ${\cal P}$- and ${\cal T}$-breaking mass terms 
\footnote{A. Kumar, S. Parlak and I. Garate (unpublished).}. 


The second main result of the present section is illustrated in  Fig. \ref{fig:twovalleyselection}, which confirms the Raman selection rule found in Sec.~\ref{sec:onevalley}. 
Under circularly polarized light, only one of the two valleys contributes to the Raman intensity [recall Eq.~(\ref{eq:sel})]; which one of the two valleys is Raman silent depends on the handedness of light.
Since the two valleys have different energy gaps at $\delta m\neq 0$, the Raman spectra for left- and right-circularly polarized light are relatively shifted in frequency. 

\begin{figure}
  \begin{center}
    \includegraphics[width=\columnwidth]{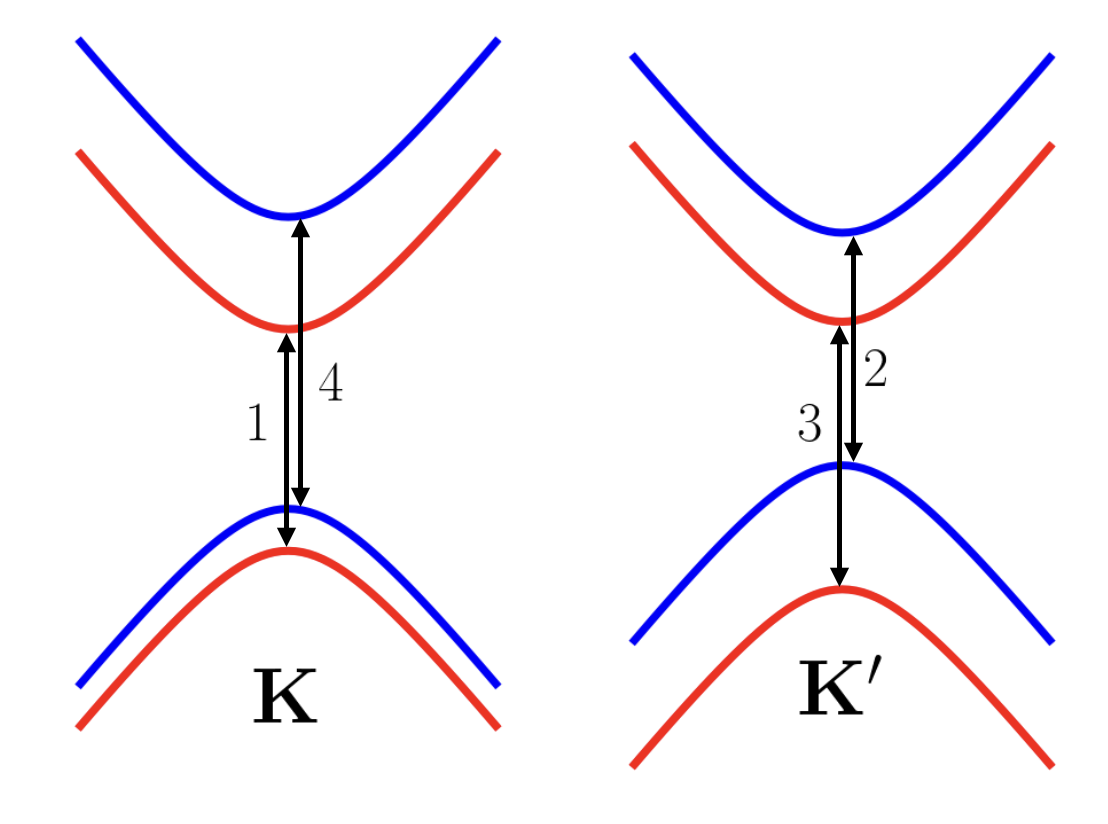}                            
\caption{Schematic low-energy electronic bands for a two-dimensional material hosting four Dirac fermions (two at each valley), adapted from Ref. \cite{qi2015giant}. 
A time-reversal-breaking perturbation renders the electronic dispersions different at the two valleys and leads to four distinct Dirac gaps, denoted with the numbers $1$ through $4$. 
The red (blue) colors denote spin-up and spin-down bands (spin-mixing terms of the Hamiltonian are neglected).
The Fermi level is assumed to be in the band gap in both valleys.
}
 \label{fig:energybands}
  \end{center}
\end{figure} 	

\begin{figure}[t]
  \begin{center}
    \includegraphics[width=\columnwidth]{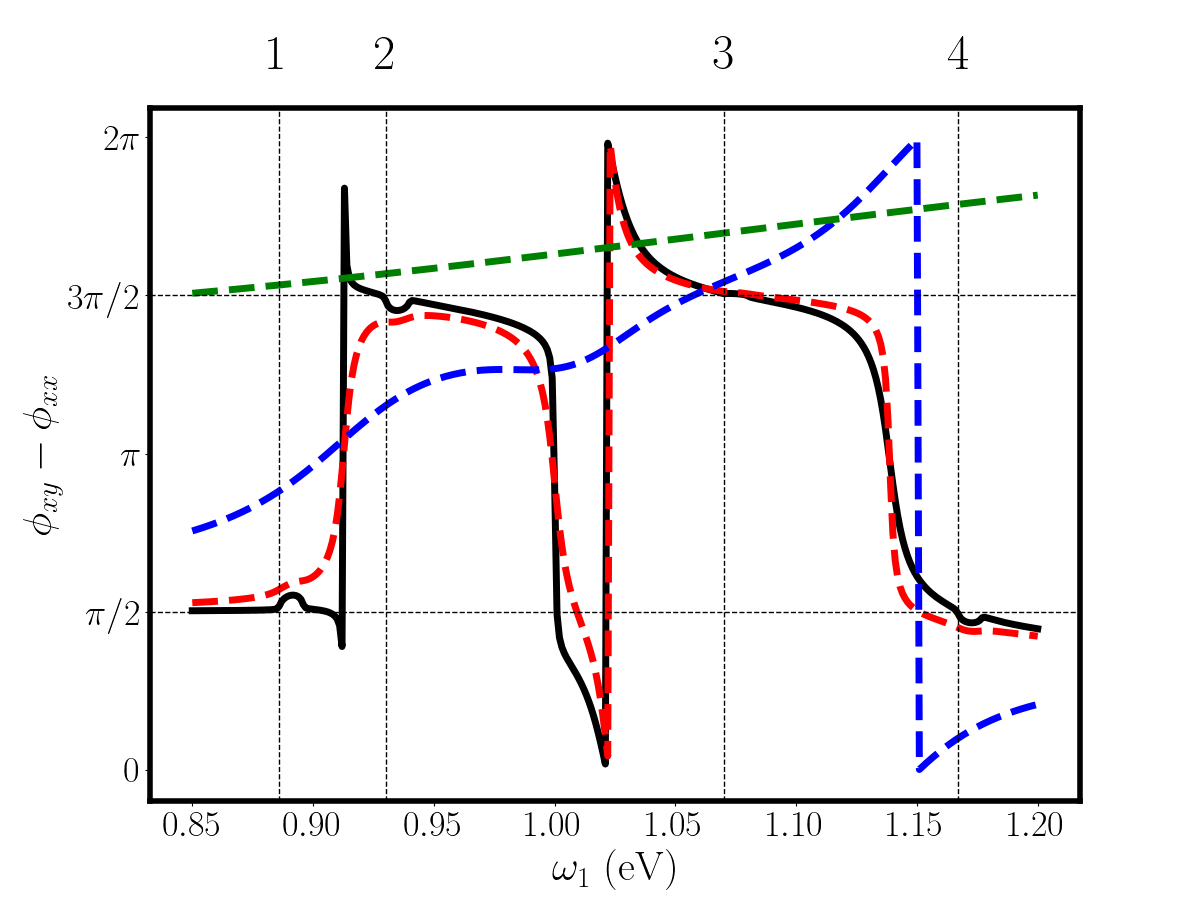}                            
\caption{Phase difference between the $R_{xx}$ and $R_{xy}$ Raman tensor elements as a function of the incident light frequency, in a low-energy model of MoTe$_2$
 with an out-of plane proximitized magnetism \cite{qi2015giant}.  The phase being defined modulo $2\pi$, jumps by multiples of $2\pi$ are visual artifacts. The dashed vertical lines denote the resonance frequencies; the numbers $1$ through $4$ correspond to the gaps in Fig. \ref{fig:energybands}. The smallest gap difference between the four Dirac fermions of Fig. \ref{fig:energybands} is $2|\delta m|\simeq 0.04 \, {\rm eV}$. The phonon energy is taken to be $\omega_0 = 0.01\, {\rm eV}$. Curves of different colors  have different values of the electronic decay rate (like in Fig. \ref{fig:twovalleyphase}).
At resonance, $\phi_{xy}-\phi_{xx}$ is quantized to $\pm \pi/2$ when $\eta\ll 2|\delta m|$.  The quantization is washed out as we increase $\eta$.} 
 \label{fig:fv}
  \end{center}
\end{figure} 

 For completeness, Fig. \ref{fig:angle_resolved} displays polar plots for the Raman intensity when the polarizations of the incident and scattered lights are $\hat {\bf e}_1=\hat{\bf e}_2=\sin \theta\hat{\bf x}+ i\cos \theta \hat{\bf y}$, with a continuously varying $\theta$; see Eq.~(\ref{eq:angle}) for the intensity expression. The electronic broadening $\eta$ is assumed to be small compared to the valley Zeeman splitting. In Fig. \ref{fig:angle_resolved}(a), the incident light frequency is significantly lower than the smallest energy gap. In that case, $|R_{xx}|$ dominates over $|R_{xy}|$ and accordingly Eq.~(\ref{eq:angle}) gives a polar plot in the form of a distorted circle. In Fig. \ref{fig:angle_resolved}(b), the incident photon frequency matches the smallest energy gap (of valley $K$). At resonance, $R_{ij}\simeq R_{i j; K}$ and, from Eq.~(\ref{eq:ratio}), it follows that $|R_{xx}|\simeq |R_{xy}|$ and $\phi_{xy}-\phi_{xx}\simeq \pi/2$. Consequently, Eq.~(\ref{eq:angle}) yields a polar plot with two lobes; the 45 degree tilt of the lobes from the horizontal axis is due to the last term in Eq.~(\ref{eq:angle}).
Fig. \ref{fig:angle_resolved}(c) displays the situation when the incident light frequency lies between the two energy gaps. As shown by Fig. \ref{fig:twovalleyratio}, in this regime $|R_{xy}|$ can become much larger than $|R_{xx}|$. Accordingly, Eq.~(\ref{eq:angle}) yields a characteristic butterfly shape to the polar plot.
In Fig. \ref{fig:angle_resolved}(d), the incident light frequency matches the largest energy gap (of valley $K'$).  At resonance, $R_{ij}\simeq R_{i j; K'}$ and, from Eq.~(\ref{eq:ratio}), it follows that $|R_{xx}|\simeq|R_{xy}|$ and $\phi_{xy}-\phi_{xx}\simeq -\pi/2$. Consequently, Eq.~(\ref{eq:angle}) yields a polar plot with two lobes; the tilt of the lobes from the horizontal axis is inverted with respect to that of Fig. \ref{fig:angle_resolved}(b), because the phase difference $\phi_{xy}-\phi_{xx}$ has opposite sign. Finally, Fig. \ref{fig:angle_resolved}(e) shows the case in which the incident light frequency significantly exceeds the largest energy gap. Like in Fig. \ref{fig:angle_resolved}(a),  $|R_{xx}|$ dominates over $|R_{xy}|$ and therefore the polar plot is predominantly circular. The rapid rotation of the intensity plot between Figs. \ref{fig:angle_resolved}(b) and \ref{fig:angle_resolved}(d) is an experimental fingerprint of the step in $\phi_{xy}-\phi_{xx}$ (cf. Fig. \ref{fig:twovalleyphase}) \footnote{ Incidentally, an anomalously large Raman rotation for an out-of-plane phonon has been recently reported in magnetized monolayer MoS$_2$, when irradiated with linearly polarized light (${\bf e}_1=(1,0,0)$ and ${\bf e}_2=(\cos\chi,\sin\chi, 0)$, with continuously varying $\chi$)
\cite{wan2021manipulating}. 
 Our theory cannot explain the observation of Ref. \cite{wan2021manipulating}, which employs a magnetic field instead of an exchange field to break time-reversal symmetry. Nevertheless, our theory does predict a significant change in the $\chi$-dependence of the Raman intensity as a function of $\delta m$, when $\omega_1$ is close to a resonant frequency.}.

 \subsection{Four Dirac fermions}
  \label{sec:four}
  
A more realistic electronic structure model of TMDC monolayers includes two Dirac fermions per valley (Fig. \ref{fig:energybands}).
Provided that we neglect the Rashba  spin-orbit interaction, the calculation of the Raman tensor for this model is an immediate extension of the one in Sec.~\ref{sec:two}; one simply needs to add the Raman tensors for all four Dirac fermions.

Figure \ref{fig:fv} shows $\phi_{xy}-\phi_{xx}$ as a function of $\omega_1$ for a low-energy model of a magnetized MoTe$_2$ monolayer \cite{qi2015giant}.
Let us first consider the case in which $\eta$ is small compared to the valley Zeeman splitting produced by the ${\cal T}$-breaking perturbation.
At resonance, $\omega_1$ matches the energy gap of one of the four Dirac fermions and thus the main contribution to the Raman tensor originates from a single Dirac fermion. Consequently, $\phi_{xy}-\phi_{xx}$ is quantized to $\pm \pi/2$ in agreement with Eq.~(\ref{eq:phase_diff}). 
The same equation also explains why $\phi_{xy}-\phi_{xx}$ changes by $\pi$ between resonances taking place at different valleys (between $1$ and $2$, or between $3$ and $4$; cf. Fig.~\ref{fig:fv}), yet it does not change between resonances taking place in the same valleys  (e.g., between $2$ and $3$ in Fig.~\ref{fig:fv}).
When $\eta$ exceeds the valley Zeeman splitting, these features are washed out.



\section{Conclusion} 
\label{sec:conc}

We have studied the amplitude for inelastic light scattering in continuum low-energy models of two-dimensional Chern insulators and magnetized monolayers of  transition metal dichalcogenides.
Specifically, we have computed the Raman tensor elements for out-of-plane lattice vibrations at zero temperature, unveiling two peculiar features that are direct consequences of the Dirac-like electronic structure.

On one hand, a selection rule emerges in the Raman tensor when the incident and scattered photons are circularly polarized.
This selection generalizes the well-known valley selection rule of optical conductivity in Dirac insulators.   
On the other hand, for an electronic model with a single Dirac fermion, the phase difference between Raman tensor elements is surprisingly quantized to $\pm \pi/2$ for any frequency of the incident light (Eq.~(\ref{eq:phase_diff})), provided that the frequency of light largely exceeds the phonon frequency.
The quantization has some robustness to perturbations (such as a nonzero electronic decay rate, an hexagonal warping, a Newtonian mass term, and variations of the ultraviolet energy cutoff) and to the presence of additional Dirac fermions insofar as the incident light frequency is less than the smallest energy gap of the insulator.

In the future, it would be interesting to attempt to experimentally observe the predicted selection rule and the phase-difference quantization. 
In transition metal dichalcogenide monolayers, a valley Zeeman splitting produced by a time-reversal-breaking perturbation is needed for our results to apply. In addition, the electronic decay rate should be smaller than the valley Zeeman splitting. These two constraints make the experimental observation of our prediction more challenging.
On the theory side, it would be desirable to test our results in real materials by using density functional theory to calculate the Raman tensor elements.


\acknowledgements
This work has been financially supported by the Natural Sciences and Engineering Research Council of Canada (Grant No. RGPIN- 2018-05385), and the Fonds de Recherche du Québec Nature et Technologies.
We acknowledge discussions with A. Khumar and A.-M. Tremblay.

\begin{widetext}


\appendix

\section{Microscopic theory expression for the Raman tensor}
\label{ap:main}
This appendix outlines the microscopic calculation of Raman tensor elements at zero temperature.  
Our starting point is the expression for the Raman tensor shown in Ref. \cite{loudon1963theory} [cf. Eq. (14) therein].
Said expression is presented in terms of many-body electronic states. 
For the purposes of our work, we are interested in a version expressed in terms of single-particle electronic states.

The matrix elements appearing in Eq. (14) of Ref. \cite{loudon1963theory} are of the type
\begin{equation}
\label{eq:me}
    \langle 0 | \hat{O}_1 | \alpha \rangle \langle \alpha | \hat{O}_2 | \beta \rangle \langle \beta | \hat{O}_3 | 0 \rangle,
\end{equation}
where  $\hat{O}_{1,2,3}$ are the interaction Hamiltonians, $| 0 \rangle$ is the many-body electronic ground state, and $| \alpha \rangle, | \beta \rangle$ are the many-body electronic excited states. 
In our case, $|0\rangle$ corresponds to a filled valence band $v$ and an empty conduction band $c$, while $|\alpha\rangle = c^\dagger_{{\bf p} c} c_{{\bf p}' v} |0\rangle$ and $|\beta\rangle = c^\dagger_{{\bf l} c} c_{{\bf l}' v} |0\rangle$ correspond to states with a single particle-hole excitation (we do not need to consider states with more than one particle-hole excitation because $\hat{O}_i$ are one-body operators in the fermionic space). Here, ${\bf p}, {\bf p}', {\bf l}$ and ${\bf l}'$ are wave vectors of the excitations. 

The states $|0\rangle$, $|\alpha\rangle$, and $|\beta\rangle$ are eigenstates of the free-electron Hamiltonian,
\begin{equation}
{\cal H}_e = \sum_{\bf k} \sum_{n\in\{c,v\}} \varepsilon_{{\bf k} n}  c^\dagger_{{\bf k}n} c_{{\bf k} n},
\end{equation}
with ${\cal H}_e |0\rangle = E_0 |0\rangle$, ${\cal H}_e |\alpha\rangle = (E_0 +\varepsilon_{{\bf p} c} - \varepsilon_{{\bf p}' v})  |\alpha\rangle$, and ${\cal H}_e |\beta\rangle = (E_0 +\varepsilon_{{\bf l} c} - \varepsilon_{{\bf l}' v})  |\beta\rangle$. Thus, $(\varepsilon_{{\bf p} c} - \varepsilon_{{\bf p}' v})$ and  $(\varepsilon_{{\bf l} c} - \varepsilon_{{\bf l}' v})$ are the excitation energies for $|\alpha\rangle$ and $|\beta\rangle$, denoted as $\omega_\alpha$ and $\omega_\beta$ in Eq.~(14) of Ref.~\cite{loudon1963theory}.

To rewrite Eq.~(\ref{eq:me}) in the single-particle formalism, we decompose
\begin{equation}
\label{eq:Oi}
    \hat{O}_i=\sum_{{\bf k},{\bf k}'}\sum_{n, n'}\langle {\bf k}n | \hat{O}_i | {\bf k}^\prime n^\prime \rangle c_{{\bf k}n}^\dagger c_{{\bf k}^\prime n^\prime},
\end{equation}
where  $| {\bf k} n \rangle$ is a single-electron energy eigenstate with eigenvalue $\varepsilon_{{\bf k} n}$ ($n\in\{c,v\}$).
Using $|\alpha\rangle = c^\dagger_{{\bf p} c} c_{{\bf p}' v} |0\rangle$, $c_{{\bf k} c} |0\rangle = \langle 0| c^\dagger_{{\bf k} c} =c^\dagger_{{\bf k} v} |0\rangle = \langle 0| c_{{\bf k} v} =0$, and the fermionic anticommutation relations, we get
\begin{equation}
 \langle 0 | \hat{O}_1 | \alpha \rangle= \langle {\bf p}' v| \hat{O}_1|{\bf p} c\rangle
\end{equation}
Similarly, using $|\beta\rangle = c^\dagger_{{\bf l} c} c_{{\bf l}' v} |0\rangle$, we have
\begin{equation}
 \langle \beta | \hat{O}_3 | 0 \rangle= \langle {\bf l} c| \hat{O}_3|{\bf l}' v\rangle.
\end{equation}
Finally, we get
\begin{equation}
\langle \alpha |\hat{O}_2 |\beta\rangle = \langle {\bf p} c|\hat{O}_2 | {\bf l} c\rangle \delta_{{\bf p}' {\bf l}'} - \langle {\bf l}' v |\hat{O}_2| {\bf p}' v\rangle \delta_{{\bf p} {\bf l}},
\end{equation}
where we have discarded the term ${\bf k}={\bf k}'$ in Eq.~(\ref{eq:Oi}).
Therefore, Eq.~(\ref{eq:me}) contains two terms; they are illustrated, for example, in Figs. 2.4(a) and 2.4(b) of Ref.~\cite{pinczuk2005fundamentals}. 


Replacing Eq.~(\ref{eq:me}) in Eq. (14) of Ref.~\cite{loudon1963theory} and neglecting finite wavevector effects, the Raman tensor for a zone-center phonon is
\begin{align}
   \frac{1}{V} \sum_{{\bf k}}& \Bigg \{ \frac{\langle {\bf k} v | {\cal H}^2_{\text{e-pt}}| {\bf k} c\rangle ({\cal H}_{\text{e-pt}}^{1,{\bf k}c,{\bf k}c}-{\cal H}_{\text{e-pt}}^{1,{\bf k}v,{\bf k}v})  \langle {\bf k} c|{\cal H}_{\text{e-pn}}|{\bf k} v\rangle}{(\varepsilon_{{\bf k}c}-\varepsilon_{{\bf k}v}+\omega_0-\omega_1-i \eta )(\varepsilon_{{\bf k}c}-\varepsilon_{{\bf k}v}+\omega_0-i \eta)} +    
   \frac{\langle {\bf k} v | {\cal H}^1_{\text{e-pt}}| {\bf k} c\rangle ({\cal H}_{\text{e-pt}}^{2,{\bf k}c,{\bf k}c}-{\cal H}_{\text{e-pt}}^{2,{\bf k}v,{\bf k}v})  \langle {\bf k} c|{\cal H}_{\text{e-pn}}|{\bf k} v\rangle}{(\varepsilon_{{\bf k}c}-\varepsilon_{{\bf k}v}+\omega_0+\omega_2-i \eta)(\varepsilon_{{\bf k}c}-\varepsilon_{{\bf k}v}+\omega_0-i \eta)} + \nonumber \\
   &\frac{\langle {\bf k} v | {\cal H}^2_{\text{e-pt}}| {\bf k} c\rangle ({\cal H}_{\text{e-pn}}^{{\bf k}c,{\bf k}c}-{\cal H}_{\text{e-pn}}^{{\bf k}v,{\bf k}v}) \langle {\bf k} c|{\cal H}^1_{\text{e-pt}}|{\bf k} v\rangle}{(\varepsilon_{{\bf k}c}-\varepsilon_{{\bf k}v}+\omega_0-\omega_1-i \eta)(\varepsilon_{{\bf k}c}-\varepsilon_{{\bf k}v}-\omega_1-i \eta)}+    
   \frac{\langle {\bf k} v | {\cal H}^1_{\text{e-pt}}| {\bf k} c\rangle ({\cal H}_{\text{e-pn}}^{{\bf k}c,{\bf k}c}-{\cal H}_{\text{e-pn}}^{{\bf k}v,{\bf k}v})  \langle {\bf k} c|{\cal H}^2_{\text{e-pt}}|{\bf k} v\rangle}{(\varepsilon_{{\bf k}c}-\varepsilon_{{\bf k}v}+\omega_0+\omega_2-i \eta)(\varepsilon_{{\bf k}c}-\varepsilon_{{\bf k}v}+\omega_2-i \eta)} +  \notag \\
   &\frac{ \langle {\bf k} v | {\cal H}_{\text{e-pn}}| {\bf k} c\rangle ({\cal H}_{\text{e-pt}}^{2,{\bf k}c,{\bf k}c}-{\cal H}_{\text{e-pt}}^{2,{\bf k}v,{\bf k}v}) \langle {\bf k} c|{\cal H}^1_{\text{e-pt}}|{\bf k} v\rangle}{(\varepsilon_{{\bf k}c}-\varepsilon_{{\bf k}v}+\omega_2-\omega_1-i \eta)(\varepsilon_{{\bf k}c}-\varepsilon_{{\bf k}v}-\omega_1-i \eta)}+    
   \frac{ \langle {\bf k} v | {\cal H}_{\text{e-pn}}| {\bf k} c\rangle ({\cal H}_{\text{e-pt}}^{1,{\bf k}c,{\bf k}c}-{\cal H}_{\text{e-pt}}^{1,{\bf k}v,{\bf k}v}) \langle {\bf k} c|{\cal H}^2_{\text{e-pt}}|{\bf k} v\rangle}{(\varepsilon_{{\bf k}c}-\varepsilon_{{\bf k}v}+\omega_2-\omega_1-i \eta)(\varepsilon_{{\bf k}c}-\varepsilon_{{\bf k}v}+\omega_2-i \eta)} 
   \Bigg \},
   \label{eq:ramanperturbation}
\end{align}
where $V$ is the sample volume, ${\cal H}_{\rm e-pt}^1$ (${\cal H}_{\rm e-pt}^2$) is the interaction between electrons and the incoming (outgoing) photon, and 
${\cal O}^{{\bf k} n,{\bf k} n'}\equiv \langle {\bf k} n |{\cal O} |{\bf k} n'\rangle$ for ${\cal O}\in\{{\cal H}_{\rm e-pt}^1, {\cal H}_{\rm e-pt}^2,{\cal H}_{\rm e-pn}\}$. 
Unlike in Ref.~\cite{loudon1963theory}, in Eq. (\ref{eq:ramanperturbation}) we have included an infinitesimal positive number $\eta$,
 necessary to correctly obtain the main results of our work.

In the language of Feynman's diagrams, Eq.~(\ref{eq:ramanperturbation}) corresponds to a time-ordered scattering amplitude with three free electron propagators and three first-order vertices (one for incoming photons, one for outgoing photons, and one for the outgoing phonon); see e.g. Figs. 2(a) and 2(b) in Ref. \cite{basko2009calculation}.
We have verified numerically that Eq.~(\ref{eq:ramanperturbation}) gives the same Raman tensor as the sum of the terms (9a) and (9b) in Ref. \cite{basko2009calculation} [these terms are associated to Figs. 2(a) and 2(b) in said article].
Specifically, the $\eta$ terms in Eq.~(\ref{eq:ramanperturbation}) are consistent with the use of time-ordered electronic Green's functions in the diagrammatic approach of  Ref. \cite{basko2009calculation}.


Reference \cite{basko2009calculation} displays additional Raman scattering amplitudes of the same order (see Fig. 2 therein), that are nevertheless not included in the treatment of Ref. \cite{loudon1963theory}.
For example, the two photon lines can coalesce into the same vertex, to make it a second-order vertex (see Fig. 2(e) of \cite{basko2009calculation}). This two-vertex diagram can be called a ``diamagnetic contribution" because it results from the second-order term in the vector potential following the application of the Peierls substitution to the electronic Hamiltonian. Explicitly, the time-ordered diamagnetic amplitude reads
\begin{equation}
\label{eq:diamagnetic}
 -\frac{1}{V} \sum_{{\bf k}}  \left[\frac{\langle {\bf k} v |V_{\rm dia}| {\bf k} c\rangle \langle {\bf k} c|{\cal H}_{\text{e-pn}}| {\bf k} v \rangle}{\omega_{{\bf k}c}-\omega_{{\bf k}v}+\omega_0-i \eta}+\frac{\langle {\bf k} c|V_{\rm dia}| {\bf k} v \rangle \langle {\bf k} v |{\cal H}_{\text{e-pn}}| {\bf k} c\rangle}{\omega_{{\bf k}c}-\omega_{{\bf k}v}-\omega_0-i \eta}\right],
\end{equation}
where $V_{\rm dia}$ is a tensor of elements $V_{\rm dia}^{ij}=\partial^2 {\cal H}_{\text{e}}/\partial k_i \partial k_j$.
This amplitude is to be added to Eq.~(\ref{eq:ramanperturbation}).
In our continuum model of massive Dirac fermions, Eq.~(\ref{eq:diamagnetic}) contributes to $R_{xx}$ when (and only when) there is a nonzero Newtonian mass term $B k^2 \sigma_z$.
There is no diamagnetic contribution to $R_{xy}$ due to the rotational symmetry of the model.
 
To be precise, the parameter $\eta$ in Eqs.~(\ref{eq:ramanperturbation}) and (\ref{eq:diamagnetic}) should be infinitesimal. That said, in various figures of the main text, we have extrapolated $\eta$ to finite values in order to phenomenologically (and roughly)  model the electronic decay rate.
 
Finally, the remaining contributions to the Raman tensor displayed in Figs. 2(c), 2(d) and 2(f) of Ref. \cite{basko2009calculation} are neglected in our case, because our electron-phonon coupling is approximated as being independent of the electronic wave vector.



\section{Analytical expressions of the Raman elements for a single Dirac fermion}
\label{ap:ratio}

In Eq. (\ref{eq:ratio}) of the main text, we have reported a simple ratio between the Raman tensor elements. The objective of this appendix is to show the analytical form of the Raman tensor elements and to justify Eq. (\ref{eq:ratio}) in the regime $\omega_1 \gg \omega_0$.
We work with the model presented in Sec.~\ref{sec:onevalley}. We neglect the hexagonal warping and the Newtonian mass terms.
From Eq.~(\ref{eq:ramanperturbation}), we find the following integrals for the Raman tensor elements:
\begin{align}
    &R_{xx}=-4  g_z  m \int_{|m|}^{\infty} d\epsilon \frac{ 8 \epsilon^4 + \omega_0^2 \omega_1 (-\omega_0+\omega_1) + 2 \epsilon^2 (\omega_0^2 + 3 \omega_0 \omega_1 - 
       3 \omega_1^2) + 
    2 m^2 (-12 \epsilon^2 + \omega_0^2 - \omega_0 \omega_1 + \omega_1^2)}{
   (2 \epsilon - \omega_0-i\eta) (2 \epsilon - \omega_1-i\eta) (2 \epsilon - \omega_1+\omega_0-i\eta) (2 \epsilon + \omega_0) (2 \epsilon + \omega_1-\omega_0) (2 \epsilon + \omega_1)} \notag\\
    &R_{xy}=i 4  g_z \int_{|m|}^{\infty} d\epsilon   \frac{(\omega_0 - 
   2 \omega_1) (\epsilon^2 (4 \epsilon^2 + \omega_0^2 + 
\omega_0 \omega_1 - \omega_1^2) + 
   m^2 (-12 \epsilon^2 + \omega_0^2 - \omega_0 \omega_1 + 
\omega_1^2))}{ (2 \epsilon - \omega_0-i\eta) (2 \epsilon - \omega_1-i\eta) (2 \epsilon - \omega_1+\omega_0-i\eta) (2 \epsilon + \omega_0) (2 \epsilon + \omega_1-\omega_0) (2 \epsilon + \omega_1) }.
\end{align}
We can readily see that, in the absence of $\eta$, $R_{xx}$ ($R_{xy}$) is purely real (imaginary) and odd (even) in mass. 
Assuming an infinitesimal $\eta$, we use the  Sokhotski–Plemelj relation $1/(x+i\eta) = p.v.(1/x) -i\pi \delta(x)$, where $p.v.$ is the principal value.
The real part of $R_{xx}$ and the imaginary part of $R_{xy}$ are
\begin{align}
&{\rm Re}(R_{xx}) =\frac{g_z m v_y}{2\omega_0 (\omega_0 - \omega_1) \omega_1 v_x}  \text{Re}\bigg [(-4 m^2 + \
\omega_0^2) \ln[-\omega_0 + 
      2 |m|] + (4 m^2 - \omega_0^2) \ln[\omega_0 + 2 |m|] +\label{eq:rerxx} \\ &
   4 m^2 \ln[-\omega_1 + 2 |m|] - 
   2 \omega_0 \omega_1 \ln[-\omega_1 + 
      2 |m|] + \omega_1^2 \ln[-\omega_1 + 2 |m|] - \nonumber \\ &
   4 m^2 \ln[\omega_0 - \omega_1 + 
      2 |m|] + \omega_0^2 \ln[\omega_0 - \omega_1 + 
      2 |m|] - \omega_1^2 \ln[\omega_0 - \omega_1 + 2 |m|] - \nonumber \\ &
   4 m^2 \ln[\omega_1 + 2 |m|] + 
   2 \omega_0 \omega_1 \ln[\omega_1 + 
      2 |m|] - \omega_1^2 \ln[\omega_1 + 2 |m|] + \nonumber \\ &
   4 m^2 \ln[-\omega_0 + \omega_1 + 
      2 |m|] - \omega_0^2 \ln[-\omega_0 + \omega_1 + 
      2 |m|] + \omega_1^2 \ln[-\omega_0 + \omega_1 + 2 |m|]) \bigg ] \nonumber\\
&{\rm Im}(R_{xy})=\frac{  g_z }{4 \omega_0 (\omega_0 - \omega_1) \omega_1} \text{Re}\bigg[(4 m^2 \
- \omega_0^2) (\omega_0 - 2 \omega_1) \ln[-\omega_0 + 
      2 |m|] \label{eq:imrxy} \\ &- (4 m^2 - \omega_0^2) (\omega_0 - 
      2 \omega_1) \ln[\omega_0 + 2 |m|] - 
   4 m^2 \omega_0 \ln[-\omega_1 + 2 |m|] + \nonumber \\ &
   8 m^2 \omega_1 \ln[-\omega_1 + 
      2 |m|] - \omega_0 \omega_1^2 \ln[-\omega_1 + 2 |m|] + 
   4 m^2 \omega_0 \ln[\omega_0 - \omega_1 + 
      2 |m|] - \omega_0^3 \ln[\omega_0 - \omega_1 + 2 |m|] -  \nonumber\\ &
   8 m^2 \omega_1 \ln[\omega_0 - \omega_1 + 2 |m|] + 
   2 \omega_0^2 \omega_1 \ln[\omega_0 - \omega_1 + 
      2 |m|] - \omega_0 \omega_1^2 \ln[\omega_0 - \omega_1 +  
      2 |m|] \nonumber\\ &+ 4 m^2 \omega_0 \ln[\omega_1 + 2 |m|] - 
   8 m^2 \omega_1 \ln[\omega_1 + 
      2 |m|] + \omega_0 \omega_1^2 \ln[\omega_1 +  
      2 |m|] \nonumber\\ &+ (-4 m^2 (\omega_0 - 
         2 \omega_1) + \omega_0 (\omega_0 - \omega_1)^2) \ln[-\
\omega_0 + \omega_1 + 2 |m|]\bigg ]\nonumber.
\end{align}
These relations are valid for all $\omega_0$ and $\omega_1$. Doing a Taylor expansion around $\omega_0=0$, we get 
 \begin{equation}
 \frac{{\rm Im}(R_{xy})}{{\rm Re}(R_{xx})}\simeq \frac{\omega_1v_y }{2 m v_x}.
 \label{ap:eq:ratio2}
 \end{equation}

Concerning the imaginary part of $R_{xx}$ and the real part of $R_{xy}$, we divide our results into five regions.

{\em Region I:} $\omega_0>2|m|$, $\omega_1-\omega_0>2|m|$, $\omega_1>2|m|$:
  
\begin{align}
&{\rm Im}(R_{xx}) = \frac{ g_z \pi  m v_x \left(2 m^2+\omega _0 (\omega_1-\omega_0)\right)}{v_y \omega_0 \omega_1 (\omega_0-\omega_1)}\label{eq:c3}  \\
      &{\rm Re}(R_{xy})= \frac{g_z \pi  \left(-2 m^2 \omega_0+4 m^2 \omega_1+\omega_0^3-2 \omega_0^2 \omega_1+\omega_0 \omega_1^2\right)}{2 \omega_0 \omega_1 (\omega_1-\omega_0)}.\nonumber
\end{align}

{\em Region II:} $\omega_0<2|m|$, $\omega_1-\omega_0>2|m|$, $\omega_1>2|m|$:
 
\begin{align}
&{\rm Im}(R_{xx}) = -\frac{ g_z \pi  m v_x (\omega_0-2 \omega_1)}{2 v_y \omega_1 (\omega_0-\omega_1)} \label{eq:c4}\\
      &{\rm Re}(R_{xy})=\frac{g_z \pi  \left(\omega_0^2-2 \omega_0 \omega_1+2 \omega_1^2\right)}{4 \omega_1 (\omega_1-\omega_0)}. \nonumber
\end{align}

{\em Region III:} $\omega_0>2|m|$, $\omega_1-\omega_0<2|m|$, $\omega_1>2|m|$:   
\begin{align}
&{\rm Im}(R_{xx}) = \frac{ g_z \pi  m v_x \left(4 m^2-\omega_0^2+\omega_1^2\right)}{2 v_y \omega_0 \omega_1 (\omega_0-\omega_1)}\label{eq:c5} \\
  &{\rm Re}(R_{xy})=\frac{g_z \pi  \left(-4 m^2 \omega_0+8 m^2 \omega_1+\omega_0^3-2 \omega_0^2 \omega_1+\omega_0 \omega_1^2\right)}{4 \omega_0 \omega_1 (\omega_1-\omega_0)}.\nonumber
\end{align}

{\em Region IV:} $\omega_0<2|m|$, $\omega_1-\omega_0<2|m|$, $\omega_1>2|m|$:
  
\begin{align}
&{\rm Im}(R_{xx}) =-\frac{ g_z \pi  m v_x \left(4 m^2+\omega_1 (\omega_1-2 \omega_0)\right)}{2 v_y \omega_0 \omega_1 (\omega_0-\omega_1)} \label{eq:c6}\\
      &{\rm Re}(R_{xy})=-\frac{ g_z \pi  \left(4 m^2 (\omega_0-2 \omega_1)+\omega_0 \omega_1^2\right)}{4 \omega_0 \omega_1 (\omega_0-\omega_1)}
      \nonumber
\end{align}

{\em Region V:} $\omega_0<2|m|$, $\omega_1-\omega_0<2|m|$, $\omega_1<2|m|$:

\begin{equation} 
{\rm Im}(R_{xx}) = {\rm Re}(R_{xy}) = 0.
\end{equation}

 
 
 When  $\omega_0\ll \omega_1$, we find
\begin{equation}
 \frac{{\rm Re}(R_{xy})}{{\rm Im}(R_{xx})}\simeq -\frac{\omega_1v_y }{2 m v_x}
  \label{ap:eq:ratio1}
 \end{equation}
for all the regions (omitting region V, which gives trivially null results). 


Combining Eqs. (\ref{ap:eq:ratio1}) and (\ref{ap:eq:ratio2}), we arrive at Eq. (\ref{eq:ratio}) in the main text. 
We have shown that this result holds for any $\omega_1$, provided that  $\omega_1 \gg \omega_0$.

\section{Analytical expression for the phase difference in the presence of a Newtonian mass term}
\label{ap:finiteB}

In the main text, we have found a simple relation [Eq. (\ref{eq:phase_diff})] for the phase difference $\phi_{xy}-\phi_{xx}$ between $R_{xy}$ and $R_{xx}$ (the $xy$ and $xx$ elements of the Raman tensor, respectively), for out-of-plane phonons. 
This relation, derived for a continuum model of a single massive Dirac fermion, surprisingly contains the ``partial" Chern number ${\rm sign}(v_x v_y m)/2$ on its right-hand side. 

In order to verify whether the appearance of the Chern number is coincidental or fundamental, we have decided to add a Newtonian mass term $B k^2 \sigma_z$ to the electronic Hamiltonian. 
When $B\neq 0$, the partial Chern number becomes
\begin{equation}
C= {\rm sign}(v_x v_y) \left({\rm sign}(m) - {\rm sign}(B)\right)/2.
\end{equation}
A natural question is whether the right-hand side of Eq. (\ref{eq:phase_diff}) is given by $\pi C/2$ when $B\neq 0$. The objective of this appendix is to answer this question in the negative.

To that end, we calculate $R_{xx}$ and $R_{xy}$ in the presence of the Newtonian mass.  We focus on the low-frequency regime ($\omega_1\ll 2 |m|$, far below resonance), where we are able to find simple analytical expressions.
For simplicity, we neglect the phonon frequency (whose effect is unimportant far from resonance) and take $v_x=v_y\equiv 1$.
Then, a convenient way to compute the Raman tensor is via
\begin{equation}
R_{ij}= \frac{\partial \Pi_{i j}}{\partial u_\lambda} =  \frac{\partial \Pi_{i j}}{\partial m} \frac{\partial m}{\partial u_\lambda} = g_z  \frac{\partial \Pi_{i j}}{\partial m},
\end{equation}
where $\Pi_{i j}$ is the polarization function.
From an explicit calculation of $\Pi_{i j}$ at small $\omega_1$ (which agrees with the result of Ref. \cite{tutschku2020momentum} available for $i=x$ and $j=y$), we arrive at
\begin{align}
\label{eq:RijB}
&R_{xx}=\frac{\omega_1^2 }{6 m^2} C+\frac{\omega_1^2 (8 B m+1) }{6 m^2 (4 B m+1)^2} C'\nonumber\\
&R_{xy}=\frac{i \omega_1^3 (-2 B m+1) }{12 m^3} C+\frac{i \omega_1^3 (6 B m+1) }{12 m^3 (4 B m+1)^2}C',
\end{align}
where
\begin{equation}
C'=  \left({\rm sign}(m) + {\rm sign}(B)\right)/2.
\end{equation}
When $B=0$, Eq.~(\ref{eq:RijB}) reduces to 
\begin{align}
&R_{xx}=\frac{\omega_1^2 }{6 m^2}\text{sign}(m) \nonumber\\
&R_{xy}=\frac{i \omega_1^3 }{12 |m|^3},
\end{align}
from which we recover Eq. (\ref{eq:ratio}) in main text.

From Eq.~(\ref{eq:RijB}), we can readily extract $\phi_{xy}-\phi_{xx}$.
Because $R_{xx}$ and $R_{xy}$ are purely real and purely imaginary, respectively, the phase difference is $\pm \pi/2$ (modulo $2\pi$).
The results are summarized in Table \ref{tab:table}.
Clearly, $\phi_{xy}-\phi_{xy}$ is not determined by the Chern number. 
Instead, $\phi_{xy}-\phi_{xy}$ in the low-frequency regime is only sensitive to the sign of the Dirac mass $m$ and is given by Eq.~(\ref{eq:phase_diff}) even when $B\neq 0$. 

\begin{table}[H]
  \begin{center}
    \begin{tabular}{|c|c|c|c|c|}
    \hline
	  & ${\rm Re}(R_{xx})$ & ${\rm Im}(R_{xy})$ & $C$ & $\phi_{xy}-\phi_{xx}$\\
          \hline
      	$m>0, B>0$ & $+$ & $+$ & $0$ & $\pi/2$  \\
       	$m<0, B>0$ & $-$ & $+$ & $-1$ & $-\pi/2$ \\
       	$m>0, B<0$ & $+$ & $+$ & $1$ & $\pi/2$ \\
          $m<0, B<0$ & $-$ & $+$ & $0$ & $-\pi/2$ \\
                  \hline              
    \end{tabular}
     \caption{The signs of the real part of $R_{xx}$ and the imaginary part of $R_{xy}$, obtained from Eq.~(\ref{eq:RijB}) in the low-frequency regime ($\omega_1\ll 2 |m|$, with a negligible phonon frequency). Also shown are the Chern number $C$ and the phase difference $\phi_{xy}-\phi_{xx}$. \label{tab:table}}       
  \end{center}
\end{table}

\end{widetext}
\bibliography{bibdirac}
\end{document}